\documentclass{l4dc2027}
\usepackage{amsmath}
\usepackage{amssymb}
\usepackage{booktabs}
\usepackage{graphicx}
\usepackage{capt-of}
\usepackage{url}
\usepackage{algorithm2e}
\usepackage{enumitem}
\usepackage{multirow}
\usepackage{graphicx}
\usepackage{placeins}
\usepackage{tikz}
\usetikzlibrary{arrows.meta}
\usepackage{pifont} 
\newtheorem{adptheorem}{Theorem}
\newtheorem{assumption}{Assumption}
\DeclareMathOperator*{\argmax}{arg\,max}

\definecolor{cOut}{HTML}{C9A93F}   
\definecolor{cIn}{HTML}{BE6464}    
\definecolor{cAcc}{HTML}{F05A00}   
\DeclareRobustCommand{\legline}[1]{\tikz[baseline=-0.6ex]\draw[#1, line width=1.8pt] (0,0) -- (4.5mm,0);}

\title[Contract2Reach]{Contract2Reach: Discretization-Induced Contractive Dynamic Programming for High-Dimensional Hamilton-Jacobi Reachability}
\usepackage{times}

\author{\Name{Jaeyoung Lee} \Email{dlwodud0206@snu.ac.kr}\\
  \addr Seoul National University
  \AND
  \Name{Donggun Lee} \Email{dlee48@ncsu.edu}\\
  \addr North Carolina State University
  \AND
  \Name{Jason J. Choi} \Email{jjhchoi@ucla.edu}\\
  \addr University of California, Los Angeles
  \AND
  \Name{Namhoon Cho} \Email{namhoon.cho@snu.ac.kr}\\
  \addr Seoul National University
}
\LinesNumbered
\DontPrintSemicolon
\begin{document}

\setlength{\textfloatsep}{2pt}
\setlength{\intextsep}{2pt}
\setlength{\floatsep}{6pt}
\maketitle

\begin{abstract}
Physics-informed neural networks (PINNs) enable undiscounted Hamilton--Jacobi (HJ) reachability analysis for high-dimensional systems, but their performance can be sensitive to hyperparameter choices and the underlying system dynamics. We introduce Contract2Reach, which recasts undiscounted finite-horizon reachability as a sequence of contractive fixed-point problems solved by backward time marching. Our formulation combines discretized temporal and spatial derivatives to construct a contractive mapping without introducing a discount factor. Under standard assumptions, we prove that the exact scheme converges to the unique viscosity solution as both the temporal and spatial step sizes tend to zero. Unlike PINNs, our training fits rollout-based value targets via a single regression, without backpropagating through spatial value gradients. In the neural implementation, adaptive zero-level-set sampling and conservative target combination address high-dimensional sample sparsity and false positives, respectively. Experiments on a parameterized vertical drone, a 13D quadrotor, and a 40D publisher--subscriber system show competitive performance against PINNs-based baselines. Boundary evaluations on the vertical drone and publisher--subscriber systems demonstrate improved fidelity and reduced learning-rate sensitivity, revealing errors that volume-based metrics can miss.
\end{abstract}


\section{Introduction}

With the need for formal safety assurances, safety-critical control has received increasing attention in recent years~\citep{bansal2017hamilton}. Especially, Hamilton–Jacobi reachability (HJR) analysis provides a rigorous framework for characterizing the Backward Reachable Tube (BRT), defined as the set of initial states from which safety or reachability can be guaranteed under system dynamics and control constraints~\citep{mitchell2005time}. Along with the set of states, HJR analysis identifies a control policy that ensures the system avoids unsafe regions or reaches a desired target set. In HJR analysis, finding the BRT requires computing a numerical solution to a certain Hamilton--Jacobi--Bellman variational inequality (HJB-VI). The result is a value function whose level sets define the avoid (or reach) set, and the gradients of the value function provide the optimal control strategy. Traditional methods solve the HJB-VI numerically over a dense state-space grid~\citep{bui2022optimizeddp,mitchell2005time}, which leads to the curse of dimensionality as the state dimension increases~\citep{darbon2016algorithms}. 

To overcome the curse of dimensionality, PINNs-based methods have been increasingly used for high-dimensional PDEs and reachability analysis~\citep{han2018solving,bansal2021deepreach}. However, direct application of PINNs~\citep{raissi2019physics} to HJB-VIs does not yield high-fidelity value functions, and a similar trend has also been observed in other domains such as fluid dynamics~\citep{chuang2022experience}. To overcome these limitations, recent PINNs-based reachability solvers incorporate various techniques such as MPC-guided supervision~\citep{feng2025bridging}, verification method~\citep{lin2024verification}, linear semi-supervision~\citep{sharpless2024linear}, and adaptive collocation sampling
~\citep{park2026forward}.
\begin{figure}[!t]
    \centering
    \resizebox{\linewidth}{!}{

\definecolor{cAcc}{HTML}{F28C28}\definecolor{cAccD}{HTML}{B3600F}\definecolor{cAccL}{HTML}{FDEBD6}
\definecolor{cInk}{HTML}{2B2B2B}\definecolor{cMid}{HTML}{5E5E5E}\definecolor{cMem}{HTML}{7A7A7A}
\definecolor{cFill}{HTML}{EAEAEA}\definecolor{cFillD}{HTML}{C8C8C8}\definecolor{cAux}{HTML}{9A9A9A}
\definecolor{cCell}{HTML}{F8F8F8}\definecolor{cCellE}{HTML}{B0B0B0}
\definecolor{cOut}{HTML}{3B6BA5}\definecolor{cOutD}{HTML}{24497A}\definecolor{cOutL}{HTML}{E3EEF9}\definecolor{cOutM}{HTML}{B6D0EE}
\definecolor{cIn}{HTML}{C0504D}\definecolor{cInD}{HTML}{8E2B28}\definecolor{cInL}{HTML}{FBE5E3}\definecolor{cInX}{HTML}{EDC4C2}
\providecommand{\crTxt}{\small}        
\providecommand{\crMath}{\normalsize}  
\providecommand{\crLab}{\small}        
\tikzset{
  c2r/.style={x=1.65cm, y=1.75cm, font=\crTxt, line cap=butt,
              >={Stealth[length=2.6mm, width=1.8mm]}},
  sig/.style={draw=cIn, line width=1.35pt},
  blk/.style={draw=cInD, fill=white, line width=1pt, rounded corners=1.2mm},
  frz/.style={draw=cOutD, fill=cOutL, line width=1.2pt, rounded corners=1.2mm},
  inb/.style={draw=cInD, fill=cInL, line width=1pt, rounded corners=1.2mm},
  gain/.style={draw=cAccD, fill=cAcc, line width=1.4pt},
  fuse/.style={draw=cAcc, line width=2.8pt},
  fuseout/.style={draw=cAcc, line width=2.2pt},   
  mem/.style={draw=cOut, line width=2.2pt},
  tap/.style={draw=cOut, line width=1.7pt},
  slc/.style={draw=cOutD, fill=cOutL, line width=1pt},
  xbus/.style={draw=cInX, line width=0.7pt, -{Stealth[length=1.6mm, width=1.1mm]}},
  xhop/.style={start angle=180, end angle=0, radius=1.1mm},
  xvhop/.style={start angle=-90, end angle=90, radius=1.1mm},
  vtgt/.style={draw=cIn, line width=1.6pt},
  lab/.style={font=\crLab, text=cMid},
  labO/.style={font=\crLab, text=cOutD},
  labI/.style={font=\crLab, text=cInD},
  hop/.style={start angle=180, end angle=0, radius=1.5mm},
}
\providecommand{\crdot}[2][cInk]{\fill[#1] (#2) circle[radius=0.75mm];}

{\fontfamily{lmr}\selectfont   
\begin{tikzpicture}[c2r]
\draw[draw=cCellE, fill=cCell, line width=1.1pt, rounded corners=4mm] (1.35,0.6) rectangle (12.1,6.35);
\node[cOutD] at (6.725,6.72) {\crMath outer loop: slice $i$};

\foreach \y/\t in {6.535/{V^h_{i+1}}, 6.145/{V^h_{i+2}}, 5.365/{V^h_{i+n}}}
  {\draw[slc] (0.1,\y-0.17) rectangle (1.0,\y+0.17); \node at (0.55,\y) {$\t$};}
\node[cOutD] at (0.55,5.79) {$\vdots$};
\draw[slc, fill=cOutM, line width=1.5pt] (12.25,6.365) rectangle (13.15,6.705); \node at (12.7,6.535) {$V^h_{i}$};
\foreach \y/\t in {6.145/{V^h_{i+1}}, 5.365/{V^h_{i+n-1}}}
  {\draw[slc] (12.25,\y-0.17) rectangle (13.15,\y+0.17); \node at (12.7,\y) {$\t$};}
\node[cOutD] at (12.7,5.79) {$\vdots$};

\draw[mem] (1.0,5.95) -- (10.55,5.95);
\draw[mem, -{Stealth[length=3.2mm, width=2.6mm]}] (11.85,5.95) -- (12.25,5.95);
\draw[draw=cOutD, fill=white, line width=1.2pt, rounded corners=1mm] (10.55,5.66) rectangle (11.85,6.24);
\foreach \k in {0,1,2,3}{
  \pgfmathsetmacro{\xc}{10.76+0.29*\k}
  \ifnum\k=0 \def\fc{cOutM}\else\def\fc{cOutL}\fi
  \draw[draw=cOutD, fill=\fc, line width=0.9pt] (\xc-0.11,5.81) rectangle (\xc+0.11,6.09);}

\draw[frz] (1.45,3.87) rectangle (2.75,4.53);
\node at (2.1,4.31) {\crMath$V^h_{i+1}(x_0)$};
\node[labO] at (2.1,4.03) {frozen};
\draw[tap, ->] (2.1,5.95) -- (2.1,4.53); \crdot[cOutD]{2.1,5.95}
\node[text=cInk, anchor=west] at (2.18,5.3) {$V^h_{i+1}$};

\fill[cAccL, rounded corners=2.5mm] (2.82,4.66) -- (6.32,4.66) -- (6.32,2.68) -- (2.17,2.68) -- (2.17,3.68) -- (2.82,3.68) -- cycle;
\draw[cAcc, line width=1.8pt, rounded corners=2.5mm]
  (2.82,4.66) -- (6.32,4.66) -- (6.32,2.68) -- (2.17,2.68) -- (2.17,3.68) -- (2.82,3.68) -- cycle;
\node[cAccD, anchor=south, font=\crMath\bfseries] at (4.55,4.68) {space--time fusion};
\node at (4.1,3.69) {$\beta=\dfrac{\Delta t_{\mathrm{time}}}{\Delta t_{\mathrm{time}}+\Delta t_{\mathrm{space}}}<1$};

\draw[draw=cOut, line width=1.5pt, ->] (2.75,4.2) -- (2.95,4.2);
\draw[gain] (2.95,3.9) -- (2.95,4.5) -- (3.72,4.2) -- cycle;
\node[text=white] at (3.22,4.2) {\crMath$\boldsymbol{1{-}\beta}$};
\draw[fuse, ->] (3.72,4.2) -- (5.77,4.2);
\node[cOutD, anchor=south, font=\crLab\bfseries] at (4.75,4.22) {time};

\draw[blk] (2.25,2.8) rectangle (5.4,3.4);
\node at (3.825,3.1) {$\max\limits_u V_{\mathrm{tgt}}\bigl(x_0+f(x_0,u)\,\Delta t_{\mathrm{space}}\bigr)$};
\draw[gain] (5.45,2.85) -- (5.45,3.35) -- (5.92,3.1) -- cycle;
\node[text=white] at (5.61,3.1) {\crMath$\boldsymbol\beta$};
\draw[fuse, ->] (5.92,3.1) -- (6.0,3.1) -- (6.0,3.98);
\node[cInD, font=\crLab\bfseries] at (5.63,3.62) {space};

\draw[draw=cAccD, fill=cAcc, line width=1.8pt] (6.0,4.2) circle[radius=3.8mm];
\draw[white, line width=1.8pt, line cap=round] ([xshift=-1.9mm]6.0,4.2) -- ([xshift=1.9mm]6.0,4.2)
                                              ([yshift=-1.9mm]6.0,4.2) -- ([yshift=1.9mm]6.0,4.2); 

\draw[draw=cAccD, fill=white, line width=1.4pt, rounded corners=1.2mm] (6.62,3.95) rectangle (8.12,4.45);
\node at (7.37,4.2) {\crMath$\min\{\ell(x_0),\,\cdot\,\}$};
\draw[fuseout, ->] (6.23,4.2) -- (6.62,4.2);
\draw[fuseout, ->] (8.12,4.2) -- ([xshift=-1.5mm]8.3,4.2) arc[hop] -- (9.05,4.2);
\node[anchor=south] at (8.7,4.24) {\crMath$\hat V_i$};
\draw[sig, fill=white] (9.25,4.2) circle[radius=3.3mm]; \node at (9.25,4.2) {$\min$};

\draw[blk] (6.1,4.85) rectangle (9.0,5.45);
\node at (7.55,5.15) {\crMath$\min\{\min_j \ell(x_j),\,V^h_{i+n}(x_{T_r})\}$};
\draw[tap, ->] (7.0,5.95) -- (7.0,5.45); \crdot[cOutD]{7.0,5.95}
\node[text=cInk, anchor=east] at (6.93,5.7) {$V^h_{i+1:i+n}$};
\draw[sig, ->] (9.0,5.15) -- (9.25,5.15) -- (9.25,4.39);
\node[anchor=west] at (9.32,4.62) {\crMath$\hat V^{\mathrm{DP}}_i$};

\draw[sig, ->] (9.25,4.01) -- (9.25,3.42);
\node[anchor=west] at (9.32,3.71) {\crMath$y$};
\draw[inb] (8.5,2.78) rectangle (10.05,3.42);
\node at (9.275,3.1) {fit $V_i(x_0)$ to $y$};

\draw[inb] (1.43,1.05) rectangle (2.77,1.65);
\node[align=center] at (2.1,1.35) {bisection on\\[-1pt]$\{V_{\mathrm{tgt}}=0\}$};
\draw[xbus] (2.1,1.65) -- (2.1,3.87);
\draw[xbus] (2.1,2.25) -- ([xshift=-1.1mm]4.6,2.25) arc[xhop] -- ([xshift=-1.1mm]8.3,2.25) arc[xhop] -- (8.7,2.25) -- (8.7,2.78);
\draw[xbus] (2.6,2.25) -- (2.6,2.8);
\draw[xbus] (7.37,2.25) -- (7.37,3.95);
\draw[xbus] (6.45,2.25) -- ([yshift=-1.1mm]6.45,4.2) arc[xvhop] -- (6.45,4.85);   
\foreach \x in {2.1,2.6,6.45,7.37} {\fill[cInX] (\x,2.25) circle[radius=0.5mm];}
\node[text=cInk, anchor=west] at (2.17,1.97) {$x_0$ samples};

\draw[sig, ->] (9.25,2.78) -- (9.25,1.68);
\draw[inb] (8.45,1.02) rectangle (10.05,1.68);
\node at (9.25,1.35) {\crMath$\dfrac{\kappa}{1-(1-\kappa)z^{-1}}$};
\node[labI] at (9.25,0.84) {Polyak averaging};
\draw[vtgt, ->] (8.45,1.35) -- (2.77,1.35);
\draw[vtgt, ->] (8.3,1.35) -- (8.3,4.85); \crdot[cIn]{8.3,1.35}
\draw[vtgt, ->] (4.6,1.35) -- (4.6,2.8); \crdot[cIn]{4.6,1.35}
\node[text=cInk, anchor=south] at (6.4,1.4) {\crMath$V_{\mathrm{tgt}}$};
\node[labI, anchor=north] at (6.4,1.28) {inner loop};

\draw[draw=cOutD, line width=1.1pt] (10.05,3.1) -- (10.1,3.1);
\draw[draw=cOutD, line width=0.9pt, fill=white] (10.14,3.1) circle[radius=0.65mm] (10.4,3.1) circle[radius=0.65mm];
\draw[draw=cOutD, line width=1.1pt] (10.17,3.12) -- (10.38,3.3);   
\draw[draw=cOut, line width=1.4pt, ->] (10.44,3.1) -- (10.76,3.1) -- (10.76,5.81);
\node[text=cInk, anchor=west] at (10.84,4.6) {$V^h_i$};

\end{tikzpicture}%
}}

    \label{fig:fig1}
    \definecolor{cOut}{HTML}{3B6BA5}   
    \definecolor{cIn}{HTML}{C0504D}    
    \definecolor{cAcc}{HTML}{F28C28}   
    {\centering \legline{cOut}~temporal information,\quad \legline{cIn}~spatial information,\quad \legline{cAcc}~space--time fusion\par}
    \vspace{-0.5em}
    \caption{Overview of the Contract2Reach framework.}
        \vspace{0.5em}
\end{figure}
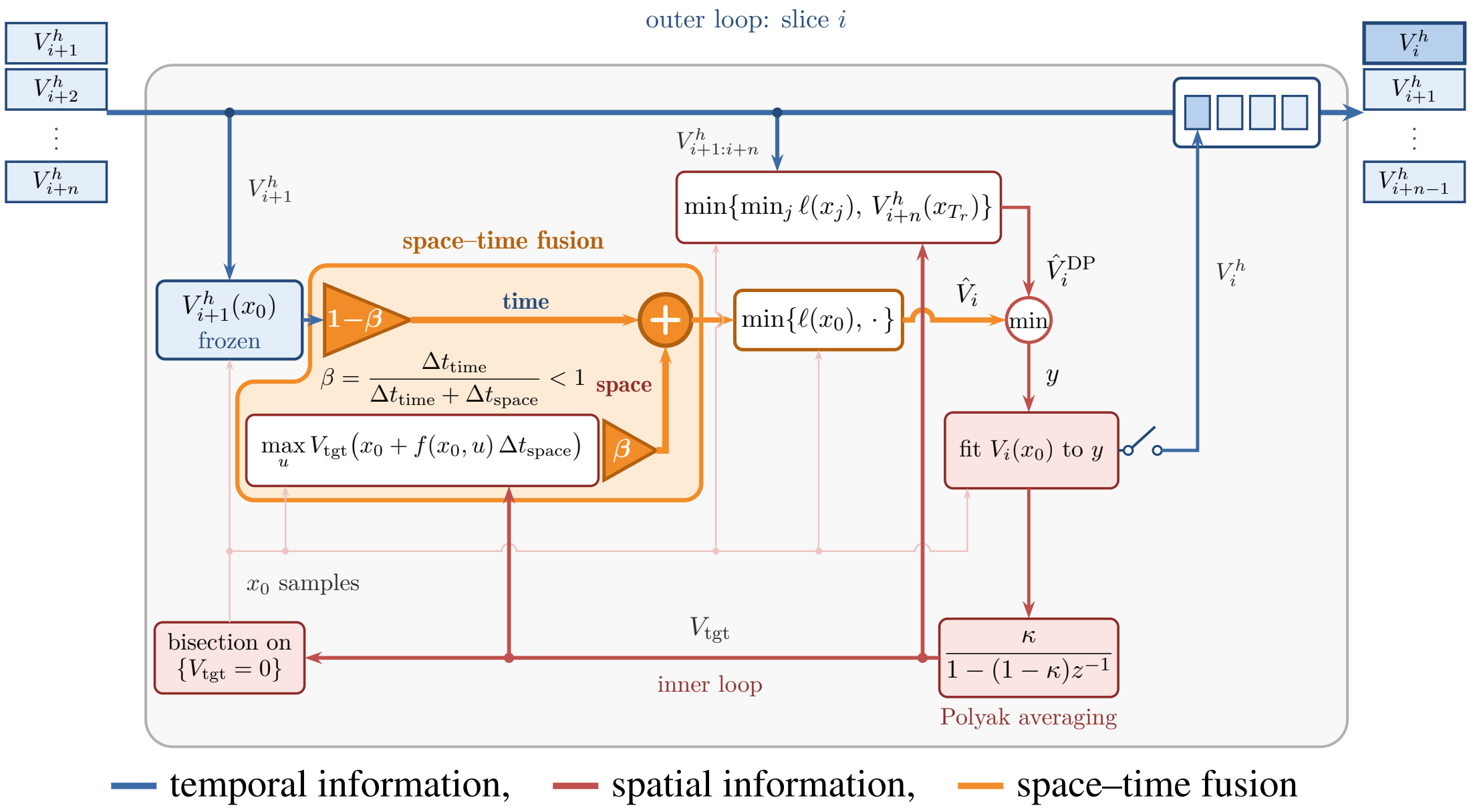
Even with these techniques, PINNs-based methods remain challenging to optimize and sensitive to training hyperparameters, particularly as the system dynamics make the underlying PDE more complex~\citep{krishnapriyan2021characterizing}. This leaves a practical gap between scalable neural HJ solvers and methods that provide stable and reliable value-function approximation without extensive hyperparameter tuning.

Dynamic programming offers a natural route to reliable learning, since it trains the value function on bootstrapped targets rather than on PDE residuals. Such training is reliable when the underlying Bellman operator is a contraction. A contraction mapping has a unique fixed point, which gives the neural network a single well-defined target to fit.
The undiscounted reachability Bellman equation is not a contraction, which is why learning-based reachability methods introduce a discount factor~\citep{fisac2019bridging}. This factor must then balance stability against accuracy. A factor strictly below one keeps the operator contractive and the training stable. A factor close to one is needed for the learned value to match the true reachability value~\citep{chen2026lambda}.

In short, PINNs-based methods directly minimize the residual of the undiscounted HJB-VI but are sensitive to training hyperparameters such as the learning rate, while DP-based methods obtain a stable learning target from a contraction but must replace the original problem with a discounted one. To address this gap, we develop a framework that combines the undiscounted formulation of PINNs-based methods with the contractive learning target of DP-based methods. Our method divides the horizon into time slices and solves the HJB-VI as a sequence of contractive fixed-point problems, starting from the terminal condition. At each time slice, Contract2Reach constructs a contraction operator by discretizing the HJB-VI and solves for its fixed point. Since the contraction arises from discretization rather than from a discount factor, the HJB-VI is left unchanged. This fixed-point structure also avoids differentiating through spatial value gradients, a key difficulty for PINNs. As a result, our method requires less hyperparameter tuning. Our contributions are as follows:

\begin{enumerate}[leftmargin=3em, topsep=6pt, itemsep=2.0pt, parsep=0pt, partopsep=0pt]
\item We introduce a \textit{discretization-induced contractive dynamic programming framework} for the original undiscounted HJB-VI, in which the contraction arises from discretization rather than from a discount factor, and prove that the resulting discrete fixed-point solution converges to the unique viscosity solution of this HJB-VI as both discretization intervals tend to zero.
\item We train the value function with a single regression objective by incorporating rollout information directly into the dynamic programming formulation. In contrast, PINNs-based methods combine boundary, HJB-VI residual, and supervision losses whose weights must be tuned, which complicates optimization.

\item By varying the learning rate, a representative optimization hyperparameter, on three benchmarks, we show that Contract2Reach recovers reachable set boundaries more accurately than PINNs-based methods, with lower boundary error and fewer topological mismatches, while keeping comparable safe set-accuracy and false-safe rates.
\end{enumerate}

\section{Other Related Work}

\noindent\textbf{Approximate DP and Continuous-Time Fitted Value Iteration.} Dynamic programming characterizes value functions as fixed points of Bellman operators, whose contraction guarantees the convergence of value iteration \citep{bertsekas1996neuro,bertsekas2018abstract}; approximate DP and fitted value iteration extend this to continuous state spaces via function approximation \citep{gordon1995stable,munos2008finite}. \citet{lutter2023continuous} apply this to continuous-time control with Continuous Fitted Value Iteration (cFVI) and Robust Fitted Value Iteration (rFVI), recovering optimal actions analytically from the value gradient. Both cFVI and rFVI rely on discounting to make the Bellman operator contractive, whereas the reachability value is defined by the minimum of $\ell$ along trajectories without discounting, so its Bellman operator is not contractive \citep{fisac2019bridging}. 
We keep the fitted-value perspective but obtain contraction from discretizing the HJB-VI and solve a fixed-point problem at each backward time slice.

\noindent\textbf{Learning for Discounted Safety and Reachability.} 
Discounted HJ methods introduce a discount factor to obtain contractive Bellman operators for learning safety values~\citep{fisac2019bridging} and to characterize invariant sets in infinite-horizon differential games~\citep{xue2022differential}. Related work develops certifiable reachability learning \citep{li2025certifiable}, neural safety certificates \citep{so2024train}, and contraction-based safety filtering \citep{kim2026deep}. \citet{solanki2026unifying} cast reachability as a discounted running-cost problem. As in \citet{fisac2019bridging}, their contraction comes from the discount, so the learned value differs from the original reachability value. In contrast, our contraction comes from discretizing the original undiscounted HJB-VI, and our fixed-point solution converges to the original value function.
\section{Preliminaries}

Consider a system of the form 
\begin{equation}
\dot{x}=f(x,u)
\end{equation}
where $f : \mathbb{R}^n\times\mathcal{U} \rightarrow \mathbb{R}^n$, with $x \in \mathbb{R}^n$ and $u \in \mathcal{U} \subset \mathbb{R}^{n_{u}}$. Here $x$ is the state, $u$ is the control. Throughout this work, we make the following assumptions on the dynamics. 
\begin{assumption}[Regularity conditions] 
\label{as:regularity}
The control set $\, \mathcal U\subset\mathbb{R}^{n_u}$ is nonempty and compact.
The dynamics $f:\mathbb{R}^n\times\mathcal U\to\mathbb{R}^n$ are continuous,
uniformly bounded, and Lipschitz continuous in the state, i.e., there exist
constants $F_f,L_f>0$ such that $\|f(x,u)\|\le F_f, \;\;
    \|f(x,u)-f(y,u)\|
    \le L_f\|x-y\|$
for all $x,y\in\mathbb{R}^n$ and $u\in\mathcal U$.
The function $\ell : \mathbb{R}^n \rightarrow \mathbb{R}$ defining $\mathcal{L}$ is bounded and Lipschitz continuous.
\end{assumption}
\subsection{Hamilton-Jacobi Reachability}
HJR analysis provides the set of initial states from which safety or reachability can be guaranteed 
under control constraints. Given a final time $T\geq0$, the BRT $\mathcal{B}$ is the set of all initial states 
from which the system enters the failure set $\mathcal{L}$ within $[0,T]$, regardless of the control taken. Formally:
\begin{equation}
\mathcal{B}=\{x:\forall u(\cdot) \in \mathcal{U},\exists\tau \in [0,T], \xi_{x,0}^{u(\cdot)}(\tau) \in \mathcal{L} \},
\end{equation}
where $\xi_{x,t}^{u(\cdot)}(\tau)$ is the state at time $\tau$ under $u(\cdot)\in\mathcal{U}$ with values in $U$, starting from $x$ at time $t$. 
We assume that the failure set $\mathcal{L}=\{x\in\mathbb{R}^n \mid \ell(x)\leq 0\}$ is the zero sublevel set of a function $\ell:\mathbb{R}^n\rightarrow\mathbb{R}$ satisfying Assumption~\ref{as:regularity}.
Then, the value function $V(x,t)=\sup_{u(\cdot)\in\mathcal{U}}\min_{s\in[t,T]}\ell(\xi_{x,t}^{u(\cdot)}(s))$ is the unique viscosity solution of the HJB-VI
\begin{equation}
\begin{cases}
  0=\min\{\ell(x)-V(x,t),D_tV(x,t)\;+H(\nabla V(x,t),x)\}, \\
  V(x,T)=\ell(x)
  \label{eq:HJR}
\end{cases}
\end{equation}
where the Hamiltonian is $H(\lambda,x)=\max_{u \in U}\lambda^\top f(x,u)$. Then, $\mathcal{B}$ and its complement, the safe set, are characterized by
\begin{equation}
V(x,0)\leq 0  \Leftrightarrow x \in \mathcal{B}, \qquad V(x,0)>0 \Leftrightarrow x \notin \mathcal{B}.
\end{equation}
Moreover, the gradient of $V(x,t)$ informs the optimal control law
\begin{equation}
u^*(x,t)=\argmax_{u \in U}\nabla V(x,t)^\top f(x,u).
\end{equation}
Under this control law, any state $x$ at time $t$ such that $V(x,t)>0$ remains outside $\mathcal{L}$ on $[t,T]$.

\subsection{Approximate Dynamic Programming}
\label{sec:adp}

We briefly review the abstract dynamic programming framework underlying approximate value and policy iteration \citep{bertsekas2018abstract}.
Let $B(X)$ denote the space of bounded real-valued functions on $X$ equipped with the sup-norm $\|J\|_{\infty}=\sup_{x\in X}|J(x)|$.
Let $U$ denote the control space, with a nonempty set $U(x)\subset U$ of feasible controls at each $x\in X$, and let $\mathcal{M}$ denote the set of admissible stationary policies, i.e., functions $\mu:X\to U$ with $\mu(x)\in U(x)$ for all $x\in X$.
Let $\mathcal{H}:X\times U\times B(X)\to\mathbb{R}$ be a given mapping.
For each $\mu\in\mathcal{M}$, let $\mathcal{T}_\mu:B(X)\to B(X)$ be the dynamic programming operator associated with policy $\mu$, defined by
\begin{equation}
  (\mathcal{T}_\mu J)(x) := \mathcal{H}\big(x,\mu(x),J\big), \qquad x\in X,\ J\in B(X).
  \label{eq:Tmu}
\end{equation}
The optimality operator $\mathcal{T}:B(X)\to B(X)$ is defined pointwise by
\begin{equation}
  (\mathcal{T}J)(x) := \inf_{\mu\in\mathcal{M}}(\mathcal{T}_\mu J)(x), \qquad x\in X,\ J\in B(X).
  \label{eq:T}
\end{equation}
We assume that $\sup_{\mu\in\mathcal{M}}\lVert\mathcal{T}_\mu J\rVert_\infty<\infty$ for every $J\in B(X)$, so that $\mathcal{T}J\in B(X)$.

\begin{definition}[Monotonicity]
A dynamic programming mapping $\mathcal{H}(x,u,J) : X\times U\times B(X)\rightarrow B(X)$ is monotone if, for any
$J,J'\in B(X)$,
\begin{equation}
    J\leq J'
    \quad\Longrightarrow\quad
    \mathcal{H}(x,u,J)\leq \mathcal{H}(x,u,J'),
    \qquad
    \forall x\in X,\; \forall u\in U(x),
    \label{eq:adp_monotonicity}
\end{equation}
where inequalities between functions are understood pointwise.
Consequently, the associated operators $\mathcal{T}_\mu$ and $\mathcal{T}$ preserve pointwise ordering.
\end{definition}

\begin{definition}[Contraction]
The dynamic programming operators are contractive if there exists
$\beta\in(0,1)$ such that, for every policy $\mu \in \mathcal{M}$ and all
$J,J'\in B(X)$,
\begin{equation}
    \|\mathcal{T}_\mu J-\mathcal{T}_\mu J'\|_\infty
    \leq
    \beta\|J-J'\|_\infty.
    \label{eq:adp_policy_contraction}
\end{equation}
\end{definition}

In approximate dynamic programming, suppose that the value update and
the policy-improvement step satisfy
\begin{align}
    \|J_{k+1}-\mathcal{T}J_k\|_\infty
    &\leq \delta,
    \label{eq:adp_value_error}\\
    \|\mathcal{T}_{\mu^k}J_k-\mathcal{T}J_k\|_\infty
    &\leq \epsilon,
    \label{eq:adp_policy_error}
\end{align}
where $\delta\geq0$ denotes the value-approximation error and
$\epsilon\geq0$ denotes the policy-improvement error.

\begin{adptheorem}[Approximate Dynamic Programming Error Bound~{\citep[Prop.~2.5.3]{bertsekas2018abstract}}]
\label{thm:adp_error_bound}
Suppose that the monotonicity and contraction conditions above hold
with contraction modulus $\beta\in(0,1)$, and that
\eqref{eq:adp_value_error}--\eqref{eq:adp_policy_error} hold uniformly
for all iterations. Then
\begin{equation}
    \limsup_{k\to\infty}
    \|J_{\mu^k}-J^\star\|_\infty
    \leq
    \frac{\epsilon+2\beta\delta}
         {(1-\beta)^2},
    \label{eq:adp_error_bound}
\end{equation}
where $J_{\mu^k} \in B(X)$ denotes the unique fixed point of
$\mathcal{T}_{\mu^k}$, and $J^\star \in B(X)$ denotes the
unique fixed point of $\mathcal{T}$.
\end{adptheorem}
In high-dimensional problems, the value function $J_k$ is commonly represented
by a neural network $J_{\theta_k}$, so that the exact dynamic programming
update is replaced by an approximate update
\begin{equation}
    J_{\theta_{k+1}} \approx \mathcal{T}_{\mu^{k}}J_{\theta_k}.
\end{equation}
The error $\delta$ captures discrepancies due to function approximation and
numerical optimization. The policy is obtained approximately from the learned
value function through greedy Hamiltonian optimization, as in continuous-time
fitted value iteration \citep{lutter2023continuous}, and the corresponding
policy-improvement error is denoted by $\epsilon$. This approximate
value-update and greedy policy-improvement structure parallels generalized
policy iteration in reinforcement learning. Theorem \ref{thm:adp_error_bound} characterizes how these
value and policy approximation errors affect the resulting policy value.

\section{Discretization-Induced Contractive Time Marching Learning for HJR }\label{sec:formulation}
We propose a contractive dynamic programming framework for high-dimensional HJ
reachability. PINNs-based methods minimize PDE residuals that depend on spatial
value gradients, requiring backpropagation through these gradients and
introducing higher-order differentiation into the computational graph. In
contrast, our method reformulates the HJB-VI as a sequence of fixed-point
problems whose training objectives depend only on value evaluations along
one-step state transitions. Spatial gradients are used only for policy recovery,
not for value-network training, resulting in a simpler optimization problem and
empirically reduced sensitivity to hyperparameter choices.
\subsection{Discretization-Induced Contraction Operator for HJR}
This section provides an overview of how the \textit{discretization-induced contractive time-marching formulation} is derived. Starting from the variational inequality in (\ref{eq:HJR}), we approximate the temporal derivative as $D_tV(x,t)\approx \frac{V(x,t+\Delta t_{\mathrm{time}})-V(x,t)}{\Delta t_{\mathrm{time}}}$ and the Hamiltonian term as $H(\nabla V(x,t),x)\approx \max_{u\in\mathcal U}\frac{V(x+f(x,u)\Delta t_{\mathrm{space}},t)-V(x,t)}{\Delta t_{\mathrm{space}}}$, i.e., discretization of directional derivative. Substituting these approximations into (\ref{eq:HJR}) yields the discrete relation used in our formulation.
\begin{align}
V(x,t)
&=
\min\left\{
\ell(x),
\frac{\Delta t_{\mathrm{space}}}
{\Delta t_{\mathrm{space}}+\Delta t_{\mathrm{time}}}
V(x,t+\Delta t_{\mathrm{time}})
\right.
\notag\\
&\qquad\left.
\qquad\qquad+
\max_{u\in\mathcal U}
\frac{\Delta t_{\mathrm{time}}}
{\Delta t_{\mathrm{space}}+\Delta t_{\mathrm{time}}}
V(x+f(x,u)\Delta t_{\mathrm{space}},t)
\right\}
\label{eq:our_formulation}
\end{align} 
Given the terminal condition $V(x,T)=\ell(x)$, $V(x,t+\Delta t_{\mathrm{time}})$ is known during backward time marching. Thus, the problem at each time step reduces to finding a fixed point for the current time-slice value function $V(x,t)$. For a fixed $V(x,t+\Delta t_{\mathrm{time}})$, any function $W\in B(\mathbb{R}^n)$, the operator $\mathcal{T}[W]$ is defined as \vspace{-0.5em}
\begin{align}
\mathcal{T}[W](x)
&:=
\min\left\{
\ell(x),
\frac{\Delta t_{\mathrm{space}}}
{\Delta t_{\mathrm{space}}+\Delta t_{\mathrm{time}}}
V(x,t+\Delta t_{\mathrm{time}})
\right.
\notag\\
&\qquad\left.
\qquad\qquad+
\max_{u\in\mathcal U}
\frac{\Delta t_{\mathrm{time}}}
{\Delta t_{\mathrm{space}}+\Delta t_{\mathrm{time}}}
W(x+f(x,u)\Delta t_{\mathrm{space}})
\right\}
\label{eq:our_operator}
\end{align}
The value function at the current time slice is then defined as the fixed point of $\mathcal{T}$, i.e., $V(\cdot,t) = \mathcal{T}[V(\cdot,t)]$. Then, the following properties hold.
\setcounter{theorem}{1}
\begin{theorem}
\label{thm:operator_properties}
Define $\mathcal{T}[W]$ as the equation of
(\ref{eq:our_operator}). For a fixed
$V(~\cdot,t+\Delta t_{\mathrm{time}})$, the operator $\mathcal{T}[W]$
is monotone and contraction mapping under the supremum norm, with
contraction factor
\[
\beta:=
\frac{\Delta t_{\mathrm{time}}}
{\Delta t_{\mathrm{space}}+\Delta t_{\mathrm{time}}}
<1.
\]
\end{theorem}
We provide the formal proof in Appendix \ref{app:mono and cont}. Theorem \ref{thm:operator_properties} ~establishes the monotonicity and contraction properties of the proposed operator. Under the assumptions required for approximate dynamic programming, the neural approximation error with respect to the corresponding discrete fixed point is bounded as stated in Theorem \ref{thm:adp_error_bound}. 

Notably, the contraction factor also determines the relative weights of the value at the future time slice and the spatially propagated value at the current time slice. These weights follow directly from the HJ consistency condition $((1-\beta)\Delta t_{\mathrm{time}}=\beta\Delta t_{\mathrm{space}})$. For a fixed $\Delta t_{\mathrm{time}}$, decreasing $\Delta t_{\mathrm{space}}$ increases $\beta$, which places more weight on the current slice but weakens the contraction.
\begin{equation}
\mathcal{T}[W](x)=\min\Bigl\{\ell(x),\;(1-\beta)\,V(x,t+\Delta t_{\mathrm{time}})+\beta\max_{u\in\mathcal U}W\bigl(x+f(x,u)\Delta t_{\mathrm{space}}\bigr)\Bigr\}.
\label{eq:our_operator_beta}
\end{equation}

Building on these results, we show that backward time marching with $\mathcal{T}$ recovers the original value function. Let $h=(\Delta t_{\mathrm{time}},\Delta t_{\mathrm{space}})$, $T=N\Delta t_{\mathrm{time}}$ for some $N\in\mathbb{N}$, and $t_i=i\Delta t_{\mathrm{time}}$ for $i=0,\dots,N$. The slices are computed backward in time, starting from the terminal condition $V_N^h=\ell$. At each step $i=N-1,\dots,0$, the later slice $V_{i+1}^h$ is already known, so \eqref{eq:our_operator} at $t=t_i$ becomes an operator on the unknown slice alone. We define $V_i^h$ as its unique fixed point, which exists because the operator is a contraction (Theorem~\ref{thm:operator_properties}). Repeating this step yields the slices $V_N^h,V_{N-1}^h,\dots,V_0^h$ on the time grid. To compare them with the continuous value function $V(x,t)$, we hold each slice constant until the next grid time and define
\[
V^h(x,t)=V_i^h(x)\quad\text{for } t\in[t_i,t_{i+1}),\qquad V^h(x,T)=\ell(x).
\]

\begin{theorem}
\label{thm:convergence_to_viscosity}
Suppose Assumption~\ref{as:regularity} holds. Then, $V^h \to V$ locally uniformly on $\mathbb{R}^n\times[0,T]$ as $h\to(0,0)$, where $V$ is the unique viscosity solution of \eqref{eq:HJR}, i.e., the HJR value function.
\end{theorem}
    We provide the formal proof in the Appendix \ref{app:convergence}. Hence, the overall error has two parts. The discretization error vanishes as $h\to(0,0)$ by Theorem~\ref{thm:convergence_to_viscosity}, and the neural approximation error is bounded by Theorem~\ref{thm:adp_error_bound}. What remains is to keep the neural approximation error small in high dimensions. In high-dimensional spaces, samples cover the state space only sparsely, and approximation errors can misclassify unsafe states as safe. Section~\ref{sec:learning} describes how our learning algorithm addresses both difficulties.

\section{Learning Slices with Neural Networks}\label{sec:learning}
 
We now describe how Contract2Reach solves them with neural networks. 
All slices share the same network architecture, but each slice $t_i$ learns its own weights $\theta_i$, and the slices are trained one at a time, backward in time. At the terminal slice $t=T$, the network is regressed onto $\ell(x)$ on training states sampled in a system-specific way (Appendix~\ref{app:exp-details}). Once a slice is trained, its weights are frozen. The frozen slice is then used to build the training targets of the next slice to be trained (Fig.~\ref{fig:fig1}).
\subsection{Adaptive Zero-Level-Set Sampling}\label{sec:sampling}
The BRT at slice $t_i$ is determined by the zero-level set of $V(\cdot,t_i)$. Since the safe controller is computed from $\nabla V$ on this set, the value function must be most accurate there. However, uniform sampling in high dimensions yields too few samples near this set to learn the boundary accurately. Prior work on space--time formulations such as DeepReach samples along trajectories to embed information about the states that the dynamics actually visit~\citep{dallas2026boundary,park2026forward}. Our framework instead solves each slice at a fixed time, so its boundary is the most critical region.

We exploit this structure with binary-search-based sampling. States visited by recent rollouts are divided into two groups by the sign of the target network $V_{\mathrm{tgt},i}$. The bisection uses only evaluations of the target network and requires neither the dynamics nor any optimal-control solution.

\subsection{Conservative Target Combination}\label{sec:conservative}
Section~\ref{sec:sampling} concentrates samples on the boundary, but trajectory information also improves learning, as the trajectory-based methods above show. We include it through the dynamic programming principle of the HJB-VI, which \citet{lygeros2004reachability} formulates for a horizon $s=n\,\Delta t_{\mathrm{time}}$ as
\begin{equation}
V(x,t_i)=\sup_{u(\cdot)}\min\Bigl\{\min_{\tau\in[t_i,\,t_i+s]}\ell\bigl(x(\tau)\bigr),\;V\bigl(x(t_i+s),\,t_i+s\bigr)\Bigr\}.
\label{eq:dpp}
\end{equation}
Starting from $x$, we roll out the system for $n$ slices under the greedy policy of the learned value functions, record the minimum of $\ell$ along the trajectory, and read the terminal value from the frozen slice $V^h_{i+n}$. Each control is computed from the slice at that time, using frozen weights for previously trained slices and the target network for the current slice. The resulting estimate $\hat V_i^{\mathrm{DP}}$ uses no value prediction at the current slice. Since the rollout policy cannot do better than the optimal policy, $\hat V_i^{\mathrm{DP}}$ is a lower bound on the true value whenever the frozen slice is accurate. MPC-guided DeepReach~\citep{feng2025bridging} also uses rollouts, but it solves a sampling-based MPC problem at each labeled state and adds the labels as a separate loss next to the PDE residual. Our rollout instead uses a single greedy policy, and we combine its estimate with the fixed-point estimate into one regression target:
\[
V_i(x)=\min\bigl\{\hat V_i(x),\,\hat V_i^{\mathrm{DP}}(x)\bigr\}.
\]
Because $\hat V_i$ relies on the target network at the current slice, it can overestimate the value and label an unsafe state safe. Taking the minimum labels a state safe only if both estimates agree, which shifts errors toward the conservative side and reduces false positives.

\section{Experimental Results}\label{sec:experiments}
 
We evaluate Contract2Reach on a parameterized vertical drone (PVD), a 13D quadrotor (Quadrotor 13D), and a 40D publisher--subscriber system (PubSub 40D). We compare it with DeepReach (vanilla)~\citep{bansal2021deepreach}, MPC DeepReach~\citep{feng2025bridging}, and Steer2Reach~\citep{park2026forward}. We highlight two results. First, Contract2Reach robustly recovers accurate boundaries across different learning rates. Second, its volume-based accuracy is competitive with that of the strongest baseline on each benchmark. Full experimental details are provided in Appendix~\ref{app:exp-details}.

\subsection{Evaluation Metrics}
We first measure how well the learned value function captures the safe set as a whole. Let $\hat{\mathcal S}=\{x\mid V_\theta(x,0)>0\}$ be the predicted safe set and $\mathcal S^*$ the set of initial states that the controller induced by $V_\theta$ actually keeps safe. We report
\begin{equation*}
\mathrm{IoU}=|\hat{\mathcal S}\cap\mathcal S^*|\,/\,|\hat{\mathcal S}\cup\mathcal S^*|,\qquad
\text{False-safe[\%]}=100\,|\hat{\mathcal S}\setminus\mathcal S^*|\,/\,|\hat{\mathcal S}|,
\end{equation*}
estimated from $10^6$ uniformly sampled initial states and their closed-loop rollouts~\citep{park2026forward}. False-safe is the safety-critical error, since it counts predicted-safe states that actually fail.

These volume-based metrics can hide geometric and topological errors of the boundary, because few samples lie near it. Yet safety is decided at the boundary, since the controller is determined by $\nabla V_\theta$, which is normal to the boundary on the zero-level set. To measure how much the boundary is deformed, we compare the predicted boundary $\hat\Gamma_s$ with the reference boundary $\Gamma^*_s$ on $S=100$ two-dimensional slices, using the distance $d(p,\Gamma)$ from a point $p$ to the nearest point of a curve $\Gamma$. Taking the maximum of these distances would let a few stray points dominate the result. We therefore use the 95th percentile $P_{95}$ instead:
\begin{equation*}
\text{B-P95}_s=P_{95}\Bigl(\{d(p,\Gamma^*_s)\}_{p\in\hat\Gamma_s}\cup\{d(p,\hat\Gamma_s)\}_{p\in\Gamma^*_s}\Bigr).
\end{equation*}
We report Top-10 B-P95, the average of the ten largest values over the slices, to focus on where the prediction is worst, as in (Fig.~\ref{fig:fig2} (c)). A distance cannot tell whether the predicted boundary consists of the right number of curves. We therefore
report Mismatch, the percentage of slices whose predicted boundary has a different number of curves from the
reference. For example, the prediction in (Fig.~\ref{fig:fig2} (g)) has four curves while the reference has two.
We also report Missing, the percentage of slices with no predicted boundary, as in (Fig.~\ref{fig:fig2} (f)). The reference boundaries are computed with a grid-based HJ reachability solver for the PVD and with the semi-analytic solution in Appendix~\ref{app:pubsub} for PubSub 40D. No reference is available for Quadrotor 13D.
 
\begin{table}[t]
\centering
\caption{Results at the learning rate with the highest IoU for each method. FS: False-safe (\%), B-P95: Top-10 B-P95, Mis.: Mismatch (\%), Miss.: Missing (\%).}
\label{tab:high_dimensional_best_observed}
\footnotesize
\setlength{\tabcolsep}{3pt}
\renewcommand{\arraystretch}{1.2}
\begin{tabular}{lccccccccccccc}
\toprule
 & \multicolumn{5}{c}{PVD} & \multicolumn{2}{c}{Quadrotor 13D} & \multicolumn{5}{c}{PubSub 40D} \\
\cmidrule(lr){2-6}\cmidrule(lr){7-8}\cmidrule(lr){9-13}
Method & IoU & FS & B-P95 & Mis. & Miss. & IoU & FS & IoU & FS & B-P95 & Mis. & Miss. \\
\midrule
DeepReach (vanilla)          & 0.186 & 81.38 & 1.953 & 100 & \textbf{0} & 0.929 & 3.71 & 0.927 & 6.60 & 0.566 & 58 & \textbf{0} \\
MPC-DeepReach  & 0.974 & 1.41 & 0.061 & \textbf{1} & 1 & \textbf{0.998} & \textbf{0.14} & \textbf{0.983} & \textbf{0.25} & 0.424 & 72 & 9 \\
Steer2Reach    & 0.953 & 2.58 & 0.080 & 2 & 2 & 0.950 & 0.20 & 0.968 & 2.53 & 0.553 & 62 & \textbf{0} \\
Contract2Reach & \textbf{0.996} & \textbf{0.06} & \textbf{0.023} & \textbf{1} & 1 & 0.993 & 0.24 & 0.963 & 3.04 & \textbf{0.350} & \textbf{27} & \textbf{0} \\
\bottomrule
\end{tabular}
\end{table}

\begin{table}[t]
\centering
\caption{Learning-rate sensitivity for each method. Results are reported as the mean and standard deviation across different learning rates for each method. Notation as in Table~\ref{tab:high_dimensional_best_observed}.}
\label{tab:high_dimensional_lr_sensitivity}
\footnotesize
\setlength{\tabcolsep}{2.5pt}
\renewcommand{\arraystretch}{1.35}
\resizebox{\linewidth}{!}{%
\begin{tabular}{lccccccccccccc}
\toprule
 & \multicolumn{5}{c}{PVD} & \multicolumn{2}{c}{Quadrotor 13D} & \multicolumn{5}{c}{PubSub 40D} \\
\cmidrule(lr){2-6}\cmidrule(lr){7-8}\cmidrule(lr){9-13}
Method & IoU & FS & B-P95 & Mis. & Miss. & IoU & FS & IoU & FS & B-P95 & Mis. & Miss. \\
\midrule
DeepReach (vanilla)          & $0.149{\pm}0.023$ & $85.11{\pm}2.28$ & $1.950{\pm}0.003$ & 100 & \textbf{0} & $0.916{\pm}0.010$ & $5.42{\pm}1.27$ & $0.523{\pm}0.479$ & $10.34{\pm}5.89$ & $0.624{\pm}0.155$ & 66.4 & 20.2 \\
MPC-DeepReach  & $0.965{\pm}0.015$ & $1.58{\pm}0.33$ & $0.096{\pm}0.047$ & 3.4 & 3.4 & $\mathbf{0.998}{\pm}0.001$ & $\mathbf{0.17}{\pm}0.04$ & $\mathbf{0.980}{\pm}0.004$ & $\mathbf{0.32}{\pm}0.23$ & $0.524{\pm}0.148$ & 52.0 & 3.2 \\
Steer2Reach    & $0.945{\pm}0.009$ & $1.74{\pm}0.89$ & $0.175{\pm}0.108$ & 4.0 & 3.6 & $0.941{\pm}0.010$ & $0.90{\pm}0.98$ & $0.944{\pm}0.027$ & $3.73{\pm}1.25$ & $0.565{\pm}0.113$ & 67.8 & 3.0 \\
Contract2Reach & $\mathbf{0.995}{\pm}0.001$ & $\mathbf{0.10}{\pm}0.04$ & $\mathbf{0.029}{\pm}0.009$ & \textbf{1.0} & 1.0 & $0.995\pm 0.001$ & $0.18\pm 0.04$ & $0.922{\pm}0.031$ & $1.46{\pm}1.09$ & $\mathbf{0.451}{\pm}0.070$ & \textbf{36.8} & \textbf{0.0} \\
\bottomrule
\end{tabular}}
\vspace{0.5em}
\end{table}

\subsection{Main Benchmark Results}

\begin{figure}[t]
\centering
\includegraphics[width=\linewidth]{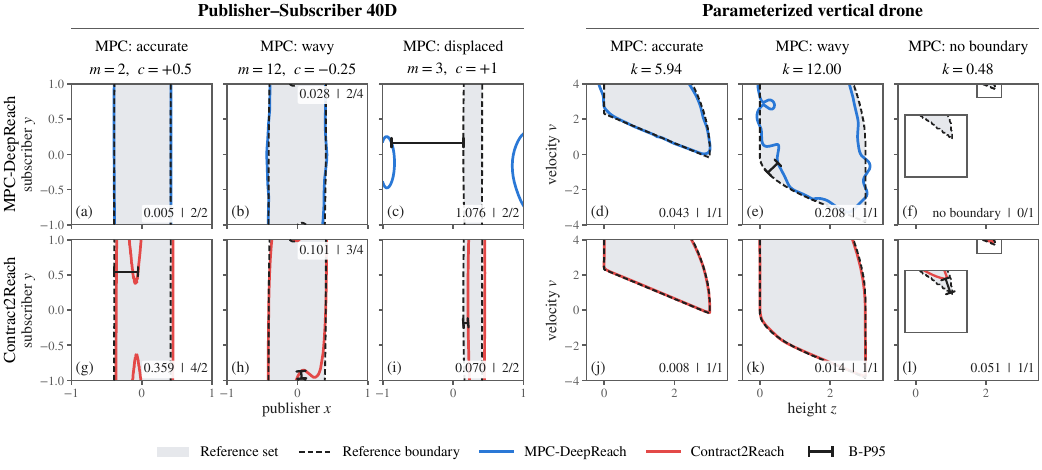}
\vspace{-1em}
\caption{Predicted boundaries of MPC-DeepReach (top) and Contract2Reach (bottom) on the same slices, each at its worst learning rate. Each PubSub slice varies the publisher state $x_0$ and the first $m$ subscriber states together, with the remaining $39-m$ subscribers fixed at $c$ (Appendix~\ref{app:pubsub}). Numbers: B-P95 $|$ predicted/reference number of boundary curves.}
\label{fig:fig2}
\end{figure}

On the PVD, Contract2Reach recovers the safe set almost exactly, with over an order of magnitude fewer false-safe states than MPC-DeepReach, and its boundary closely follows the reference on every slice. Deepreach (vanilla), in contrast, converge to a nonphysical solution that labels nearly the entire state space safe. On Quadrotor 13D, Contract2Reach is close to MPC-DeepReach on both volume-based metrics, and the small gap is expected because MPC-DeepReach includes a fine-tuning stage designed specifically to suppress false positives~\citep{feng2025bridging}. Appendix~\ref{app:result-details} provides more detailed results, including additional slices for all systems at every learning rate.

PubSub 40D illustrates why we evaluate the boundary separately. MPC-DeepReach scores best on the volume-based metrics, yet on some slices it misplaces the boundary entirely and classifies the entire slice as one class (Fig.~\ref{fig:fig2}, top). Such slices occupy a negligible fraction of the 40-dimensional state space, so these failures leave IoU and False-safe almost unchanged. Contract2Reach instead places a boundary on every slice and preserves the topology of the reference far more often, at the cost of a higher False-safe rate than MPC-DeepReach.

Across learning rates (Table~\ref{tab:high_dimensional_lr_sensitivity}), the boundaries learned by Contract2Reach change little, whereas those of the baselines vary considerably. On PubSub 40D, a poorly chosen learning rate makes Contract2Reach more conservative rather than less safe: it misses part of the safe set, but its false-safe rate does not increase, as expected from the conservative target combination (Section~\ref{sec:conservative}).

\section{Conclusion}
We presented Contract2Reach, a contractive dynamic programming framework for undiscounted finite-horizon HJ reachability. We achieve contraction through a novel discretization scheme that combines temporal and spatial derivatives without introducing a discount factor, and prove that the resulting discrete fixed-point solution converges to the unique viscosity solution of the original HJB-VI as both discretization intervals tend to zero. Neural training uses a single regression on rollout-based targets, with adaptive zero-level-set sampling and conservative target combination addressing sample sparsity and false positives, respectively. Experiments on systems with up to 40 dimensions show competitive volume-based performance, and boundary evaluations show the lowest boundary error with relatively small learning-rate sensitivity while revealing geometric and topological errors hidden by volume metrics. 

Future work includes a systematic study of how temporal and spatial discretization step sizes affect the accuracy of the learned boundary. We also plan to adapt the contraction factor to the relative uncertainty of the value estimates at the current and next slices while preserving HJ consistency.

\subsection*{Generative AI Use}

Generative AI tools (Claude, Anthropic) were used for writing and editing (including drafting the abstract) and for generating and debugging the code, including the plotting scripts for the figures and the TikZ code for Figure 1. The authors reviewed and verified all AI-assisted content and take responsibility for the final content of the paper.

\acks{This work was supported by the Strategic Hub for International Research Collaboration project of Seoul National University (SNU).
 We would like to thank Inkyu Jang for helpful discussions related to the paper.}
 



\bibliography{references}

@inproceedings{bansal2021deepreach,
  title={Deepreach: A deep learning approach to high-dimensional reachability},
  author={Bansal, Somil and Tomlin, Claire J},
  booktitle={2021 IEEE International Conference on Robotics and Automation (ICRA)},
  pages={1817--1824},
  year={2021},
  organization={IEEE}
}

@inproceedings{bansal2017hamilton,
  title={Hamilton-jacobi reachability: A brief overview and recent advances},
  author={Bansal, Somil and Chen, Mo and Herbert, Sylvia and Tomlin, Claire J},
  booktitle={2017 IEEE 56th annual conference on decision and control (CDC)},
  pages={2242--2253},
  year={2017},
  organization={IEEE}
}

@article{feng2025bridging,
  title={Bridging model predictive control and deep learning for scalable reachability analysis},
  author={Feng, Zeyuan and Qiu, Le and Bansal, Somil},
  journal={arXiv preprint arXiv:2505.03830},
  year={2025}
}

@article{sharpless2024linear,
  title={Linear supervision for nonlinear, high-dimensional neural control and differential games},
  author={Sharpless, William and Feng, Zeyuan and Bansal, Somil and Herbert, Sylvia},
  journal={arXiv preprint arXiv:2412.02033},
  year={2024}
}

@article{park2026forward,
  title={Forward Trajectory Steering for Hamilton-Jacobi Reachability Analysis},
  author={Park, Sungje and Tu, Stephen},
  journal={arXiv preprint arXiv:2608.11480},
  year={2026}
}

@article{dallas2026boundary,
  title={Boundary Sampling to Learn Predictive Safety Filters via Pontryagin's Maximum Principle},
  author={Dallas, James and Lew, Thomas and Talbot, John and DeCastro, Jonathan and Bansal, Somil and Subosits, John},
  journal={arXiv preprint arXiv:2604.13325},
  year={2026}
}

@book{bertsekas1996neuro,
  title={Neuro-dynamic programming},
  author={Bertsekas, Dimitri P and Tsitsiklis, John N and Bertsekas, Dimitri P and Tsitsiklis, John N and Bertsekas, Dimitri P and Bertsekas, Dimitri P},
  volume={5},
  year={1996},
  publisher={Athena scientific, Belmont, MA}
}

@incollection{gordon1995stable,
  title={Stable function approximation in dynamic programming},
  author={Gordon, Geoffrey J},
  booktitle={Machine learning proceedings 1995},
  pages={261--268},
  year={1995},
  publisher={Elsevier}
}

@article{munos2008finite,
  title={Finite-Time Bounds for Fitted Value Iteration.},
  author={Munos, R{\'e}mi and Szepesv{\'a}ri, Csaba},
  journal={Journal of Machine Learning Research},
  volume={9},
  number={5},
  year={2008}
}

@incollection{lutter2023continuous,
  title={Continuous-time fitted value iteration for robust policies},
  author={Lutter, Michael},
  booktitle={Inductive Biases in Machine Learning for Robotics and Control},
  pages={71--111},
  year={2023},
  publisher={Springer}
}

@inproceedings{fisac2019bridging,
  title={Bridging hamilton-jacobi safety analysis and reinforcement learning},
  author={Fisac, Jaime F and Lugovoy, Neil F and Rubies-Royo, Vicen{\c{c}} and Ghosh, Shromona and Tomlin, Claire J},
  booktitle={2019 International Conference on Robotics and Automation (ICRA)},
  pages={8550--8556},
  year={2019},
  organization={IEEE}
}

@article{solanki2026unifying,
  title={Unifying Hamilton-Jacobi Reachability and Reinforcement Learning},
  author={Solanki, Prashant and El-Hajj, Isabelle and van Beers, Jasper and van Kampen, Erik-Jan and de Visser, Coen},
  journal={arXiv preprint arXiv:2601.08050},
  year={2026}
}

@article{li2025certifiable,
  title={Certifiable reachability learning using a new lipschitz continuous value function},
  author={Li, Jingqi and Lee, Donggun and Lee, Jaewon and Dong, Kris Shengjun and Sojoudi, Somayeh and Tomlin, Claire},
  journal={IEEE Robotics and Automation Letters},
  volume={10},
  number={4},
  pages={3582--3589},
  year={2025},
  publisher={IEEE}
}

@article{kim2026deep,
  title={Deep QP Safety Filter: Model-free Learning for Reachability-based Safety Filter},
  author={Kim, Byeongjun and Kim, H Jin},
  journal={arXiv preprint arXiv:2601.21297},
  year={2026}
}

@inproceedings{so2024train,
  title={How to train your neural control barrier function: Learning safety filters for complex input-constrained systems},
  author={So, Oswin and Serlin, Zachary and Mann, Makai and Gonzales, Jake and Rutledge, Kwesi and Roy, Nicholas and Fan, Chuchu},
  booktitle={2024 IEEE International Conference on Robotics and Automation (ICRA)},
  pages={11532--11539},
  year={2024},
  organization={IEEE}
}

@article{lygeros2004reachability,
  title={On reachability and minimum cost optimal control},
  author={Lygeros, John},
  journal={Automatica},
  volume={40},
  number={6},
  pages={917--927},
  year={2004},
  publisher={Elsevier}
}

@article{mitchell2005time,
  title={A time-dependent Hamilton-Jacobi formulation of reachable sets for continuous dynamic games},
  author={Mitchell, Ian M and Bayen, Alexandre M and Tomlin, Claire J},
  journal={IEEE Transactions on automatic control},
  volume={50},
  number={7},
  pages={947--957},
  year={2005},
  publisher={IEEE}
}

@article{barles1991convergence,
  title={Convergence of approximation schemes for fully nonlinear second order equations},
  author={Barles, Guy and Souganidis, Panagiotis E},
  journal={Asymptotic analysis},
  volume={4},
  number={3},
  pages={271--283},
  year={1991},
  publisher={SAGE Publications, Sage UK: London, England}
}

@article{hirsch2026extending,
  title={Extending and Unifying the Fundamental Tasks of Hamilton-Jacobi Reachability Analysis},
  author={Hirsch, Dylan and Sharpless, William and Lee, Donggun and Herbert, Sylvia},
  journal={arXiv preprint arXiv:2608.18060},
  year={2026}
}

@book{kreyszig1991introductory,
  title={Introductory functional analysis with applications},
  author={Kreyszig, Erwin},
  year={1991},
  publisher={John Wiley \& Sons}
}

@article{raissi2019physics,
  title={Physics-informed neural networks: A deep learning framework for solving forward and inverse problems involving nonlinear partial differential equations},
  author={Raissi, Maziar and Perdikaris, Paris and Karniadakis, George E},
  journal={Journal of Computational physics},
  volume={378},
  pages={686--707},
  year={2019},
  publisher={Elsevier}
}

@article{bui2022optimizeddp,
  title={Optimizeddp: An efficient, user-friendly library for optimal control and dynamic programming},
  author={Bui, Minh and Hu, Hanyang and He, Chong and Lu, Michael and Giovanis, George and Shriraman, Arrvindh and Chen, Mo},
  journal={arXiv preprint arXiv:2204.05520},
  year={2022}
}

@article{han2018solving,
  title={Solving high-dimensional partial differential equations using deep learning},
  author={Han, Jiequn and Jentzen, Arnulf and E, Weinan},
  journal={Proceedings of the National Academy of Sciences},
  volume={115},
  number={34},
  pages={8505--8510},
  year={2018},
  publisher={National Academy of Sciences}
}

@inproceedings{lin2024verification,
  title={Verification of neural reachable tubes via scenario optimization and conformal prediction},
  author={Lin, Albert and Bansal, Somil},
  booktitle={6th Annual Learning for Dynamics \& Control Conference},
  pages={719--731},
  year={2024},
  organization={PMLR}
}

@article{darbon2016algorithms,
  title={Algorithms for overcoming the curse of dimensionality for certain Hamilton--Jacobi equations arising in control theory and elsewhere},
  author={Darbon, J{\'e}r{\^o}me and Osher, Stanley},
  journal={Research in the Mathematical Sciences},
  volume={3},
  number={1},
  pages={19},
  year={2016},
  publisher={Springer}
}

@article{krishnapriyan2021characterizing,
  title={Characterizing possible failure modes in physics-informed neural networks},
  author={Krishnapriyan, Aditi and Gholami, Amir and Zhe, Shandian and Kirby, Robert and Mahoney, Michael},
  journal={Advances in neural information processing systems},
  volume={34},
  pages={26548--26560},
  year={2021}
}

@inproceedings{xue2022differential,
  title={Differential games based on invariant sets generation},
  author={Xue, Bai and Wang, Qiuye and Zhan, Naijun and Fr{\"a}nzle, Martin and Feng, Shenghua},
  booktitle={2022 American Control Conference (ACC)},
  pages={1285--1292},
  year={2022},
  organization={IEEE}
}

@article{chen2026lambda,
  title   = {{$\lambda$}-Reachability: Geometric-Horizon Safety {B}ellman Equations for Humanoid Safety},
  author  = {Chen, Rui and Li, Shangtao and Sun, Yifan and Liu, Changliu},
  journal = {arXiv preprint arXiv:2606.16022},
  year    = {2026}
}

@article{chuang2022experience,
  title={Experience report of physics-informed neural networks in fluid simulations: pitfalls and frustration},
  author={Chuang, Pi-Yueh and Barba, Lorena A},
  journal={arXiv preprint arXiv:2205.14249},
  year={2022}
}

@book{bertsekas2018abstract,
  title={Abstract dynamic programming},
  author={Bertsekas, Dimitri P},
  year={2018},
  publisher={Athena Scientific, Nashua, NH, USA}
}

\newpage
\appendix
\section{Contract2Reach Details and Proofs}
\subsection{Proof of Theorem 2}
\label{app:mono and cont}

\begin{proof}
Let
\begin{equation}
     \alpha
    :=
    \frac{\Delta t_{\mathrm{space}}}
    {\Delta t_{\mathrm{space}}
    +\Delta t_{\mathrm{time}}},
    \qquad
    \beta
    :=
    \frac{\Delta t_{\mathrm{time}}}
    {\Delta t_{\mathrm{space}}
    +\Delta t_{\mathrm{time}}},
\end{equation}
so that
\[
    \alpha+\beta=1,
    \qquad
    0<\beta<1.
\]
For fixed $\bar V(x):=V(x,t+\Delta t_{time})$, write
\begin{equation}
    \mathcal{T}[W](x)=
    \min\left\{
    \ell(x),
    \alpha\bar V(x)
    +
    \beta\mathcal{M}[W](x)
    \right\},
\end{equation}
where
\[
\mathcal M[W](x)
:=
\max_{u\in\mathcal U}
W\!\left(
x+\Delta t_{\mathrm{space}}f(x,u)
\right).
\]

\paragraph{Monotonicity.}
Suppose that $W_{1}(x)\le W_{2}(x)$ for all $x$. Then, for every
$u\in\mathcal U$,
\begin{equation}
W_{1}\!\left(
x+\Delta t_{\mathrm{space}}f(x,u)
\right)
\le
W_{2}\!\left(
x+\Delta t_{\mathrm{space}}f(x,u)
\right).    
\end{equation}
Taking the maximum over $u$ preserves the inequality, and therefore
\[
    \mathcal M[W_{1}](x)
    \le
    \mathcal M[W_{2}](x).
\]
Since $\beta>0$ and the map $z\mapsto\min\{\ell(x),z\}$ is nondecreasing,
\begin{equation}
    \mathcal{T}[W_{1}](x)
    \le
    \mathcal{T}[W_{2}](x)
\end{equation}
for every $x$. Hence $\mathcal{T}$ is monotone.

\paragraph{Contraction.}
For any bounded functions $W_{1}$ and $W_{2}$,
\begin{equation}
\begin{aligned}
|\mathcal M[W_{1}](x)-\mathcal M[W_{2}](x)|
&=
\left|
\max_{u\in\mathcal U}
W_{1}(x+\Delta t_{\mathrm{space}}f(x,u))
-
\max_{u\in\mathcal U}
W_{2}(x+\Delta t_{\mathrm{space}}f(x,u))
\right|
\\
&\le
\max_{u\in\mathcal U}
\left|
W_{1}(x+\Delta t_{\mathrm{space}}f(x,u))
-
W_{2}(x+\Delta t_{\mathrm{space}}f(x,u))
\right|
\\
&\le
\|W_{1}-W_{2}\|_\infty.
\end{aligned}
\end{equation}
Thus $\mathcal M$ is nonexpansive under the supremum norm.

Next, the scalar map $z\mapsto\min\{\ell(x),z\}$ is $1$-Lipschitz. Therefore,
\begin{equation}
\begin{aligned}
|\mathcal{T}[W_{1}](x)-\mathcal{T}[W_{2}](x)|
&\le
\left|
\alpha\bar V(x)
+\beta\mathcal M[W_{1}](x)
-
\alpha\bar V(x)
-\beta\mathcal M[W_{2}](x)
\right|
\\
&=
\beta
|\mathcal M[W_{1}](x)-\mathcal M[W_{2}](x)|
\\
&\le
\beta\|W_{1}-W_{2}\|_\infty.
\end{aligned}
\end{equation}
Taking the supremum over $x$ gives
\begin{equation}
    \|\mathcal{T}[W_{1}]-\mathcal{T}[W_{2}]\|_\infty
    \le
    \beta\|W_{1}-W_{2}\|_\infty.
    \label{eq:contraction}
\end{equation}
\end{proof}
Since $\mathcal{T}$ is a contraction operator by Theorem~\ref{thm:operator_properties},
the Banach fixed-point theorem guarantees that, under exact operator
evaluation, $\mathcal{T}$ admits a unique fixed point $V^\star$. Moreover, for
any bounded initialization $V^{(0)}$, the iteration $V^{(k+1)} = \mathcal{T}[V^{(k)}]$
converges to $V^\star$ in the supremum norm.

\subsection{Proof of Theorem 3}
\label{app:convergence}
This section proves Theorem~\ref{thm:convergence_to_viscosity}.
The argument combines two standard ingredients from the
Hamilton--Jacobi literature.

First, under standard regularity conditions, finite-horizon
HJ reachability value functions are characterized as viscosity
solutions of the corresponding HJ terminal-value problems, with
uniqueness following from an appropriate comparison principle
\citep{lygeros2004reachability,mitchell2005time}.
Recent HJR theory follows the same viscosity-solution framework for
more general reachability formulations and weaker technical
assumptions \citep{hirsch2026extending}. For the specific HJ variational inequality considered here, we verify the comparison principle required by our convergence argument directly in Step~5 under Assumption~\ref{as:regularity}.

Second, to establish convergence of the proposed discretization, we
use the monotone approximation framework of
\citep{barles1991convergence}. We recall below the form of their
convergence result needed in our setting.

\begin{theorem}[Barles--Souganidis convergence theorem, specialized form]
\label{thm:barles_souganidis_specialized}
Consider a family of discrete approximation schemes
\[
    S_h(z,u_h(z),u_h)=0,
    \qquad h\to0,
\]
for a continuous first-order limiting Hamilton--Jacobi variational inequality
\[
    F(z,u(z),Du(z))=0.
\]
Suppose that:

\begin{enumerate}
    \item the scheme is \emph{monotone}: for any bounded functions
    $\psi\ge\widetilde\psi$,
    \[
        S_h(z,r,\psi)
        \le
        S_h(z,r,\widetilde\psi);
    \]

    \item the scheme is \emph{stable}: its discrete solutions
    $u_h$ are uniformly bounded independently of $h$;

    \item the normalized scheme residual is \emph{consistent}: for every smooth bounded test
    function $\phi$, every sequence $h_k\to0$ and discrete nodes $z_k\to z$ with $z$ in the PDE interior, and every
    $\xi_k\to0$, 
    \[
        S_{h_{k}}(z_k,\phi(z_k)+\xi_k,\phi+\xi_k)\to F(z,\phi(z),D \phi(z));
    \]

    \item the limiting equation satisfies a comparison between bounded upper-semicontinuous viscosity subsolutions and bounded lower-semicontinuous viscosity supersolutions with ordered boundary traces.
\end{enumerate}

Assume also that both half-relaxed limits of the discrete solutions attain the prescribed boundary data. Then uh converge locally uniformly to the unique continuous viscosity solution in the PDE interior and at the boundary where this trace condition holds. For time-discrete schemes, convergence is understood through discrete nodes, or through a time extension with the same half-relaxed limits.
\end{theorem}

The theorem above is a specialization of the monotone-scheme
convergence result of \citep{barles1991convergence} to the
first-order terminal-value problem considered here, with normalization absorbed into $S_h$ and terminal traces verified separately. In the original framework, consistency is formulated through separate $\limsup$ and $\liminf$ inequalities involving the upper and lower
semicontinuous envelopes $F^*$ and $F_*$. In our setting, the limiting
HJ operator $F$ is continuous, so $F^*=F_*=F$. Since we directly prove
that the scheme residual converges to $F$ for every sequence
$h_k\to0$, $z_k\to z$, and $\xi_k\to0$, both consistency inequalities
of the original theorem follow immediately.

Accordingly, the remainder of the proof verifies, in order,
well-posedness and stability of the discrete solution, consistency
with the terminal condition, monotonicity of the scheme residual,
and consistency with the limiting HJ variational inequality. We then
establish the required comparison principle directly in Step 5 and invoke Theorem~\ref{thm:barles_souganidis_specialized}.

\begin{proof}
The case $T=0$ is immediate from the terminal condition. For $T>0$, let
\[
0=t_0<t_1<\cdots<t_N=T,
\qquad
t_{i+1}-t_i=\Delta t_{\mathrm{time}},
\]
where, for simplicity, $T=N\Delta t_{\mathrm{time}}$, and let $\Delta t_{\mathrm{space}}>0$
denote the propagation interval appearing in
\eqref{eq:our_formulation}. Define
\begin{equation}
\begin{aligned}
a_h&:=\frac{\Delta t_{\mathrm{space}}}
 {\Delta t_{\mathrm{time}}+\Delta t_{\mathrm{space}}},
& b_h&:=\frac{\Delta t_{\mathrm{time}}}
 {\Delta t_{\mathrm{time}}+\Delta t_{\mathrm{space}}},\\
\lambda_h&:=\frac{\Delta t_{\mathrm{time}}\Delta t_{\mathrm{space}}}
 {\Delta t_{\mathrm{time}}+\Delta t_{\mathrm{space}}}.
\end{aligned}
\end{equation}
Here $h=(\Delta t_{\mathrm{time}},\Delta t_{\mathrm{space}})$,
$a_h=1-b_h$, and $b_h=\beta$ in Theorem~\ref{thm:operator_properties}.
Notice that
\[
a_h+b_h=1,
\qquad
a_h\Delta t_{\mathrm{time}}=b_h\Delta t_{\mathrm{space}}=\lambda_h.
\]

Let $B(\mathbb R^n)$ denote the Banach space of bounded
real-valued functions equipped with $\|\cdot\|_\infty$, as in the main
text. For $v\in B(\mathbb R^n)$, define
\begin{equation}
\mathcal M_{\Delta t_{\mathrm{space}}}[v](x)
:=
\sup_{u\in\mathcal U}
v\!\left(x+\Delta t_{\mathrm{space}} f(x,u)\right).
\end{equation}
The supremum is well defined for every bounded $v$. When $v$ is
continuous, it is attained by compactness of $\mathcal U$ and
continuity of $f$, and hence agrees with the maximum in the main text.
The exact discrete scheme associated with
\eqref{eq:our_formulation} can then be written as
\begin{equation}
V_i^h(x)
=
\min\left\{
\ell(x),
a_hV_{i+1}^h(x)
+
b_h\mathcal M_{\Delta t_{\mathrm{space}}}[V_i^h](x)
\right\},
\label{eq:app_scheme}
\end{equation}
with terminal condition
\begin{equation}
V_N^h(x)=\ell(x).
\label{eq:app_terminal}
\end{equation}
As in the main text, we compare different time grids through the piecewise-constant extension
\begin{equation}
V^h(x,t)=
\begin{cases}
V_i^h(x),&t_i\le t<t_{i+1},\\
\ell(x),&t=T,
\end{cases}
\label{eq:app_time_extension}
\end{equation}
which is the function $V^h$ in Theorem~\ref{thm:convergence_to_viscosity}.

\paragraph{Step 1: Well-posedness and uniform stability.}

For fixed $w\in B(\mathbb R^n)$, define the single-slice operator
\begin{equation}
\mathcal T_h^w[v](x)
:=
\min\left\{
\ell(x),
a_hw(x)
+
b_h\mathcal M_{\Delta t_{\mathrm{space}}}[v](x)
\right\}.
\end{equation}

This operator maps $B(\mathbb R^n)$ into itself. By the contraction inequality established in Theorem~\ref{thm:operator_properties}, we have,
for any bounded $v_1,v_2$,
\begin{equation}
\left\|
\mathcal T_h^w[v_1]
-
\mathcal T_h^w[v_2]
\right\|_\infty
\le
b_h\|v_1-v_2\|_\infty.
\end{equation}
Because $0<b_h<1$, $\mathcal T_h^w$ is a contraction.
Hence, by the Banach fixed-point theorem \citep{kreyszig1991introductory}, each time slice
\eqref{eq:app_scheme} admits a unique solution $V_i^h$.
This is precisely the fixed-point property established in
Theorem~\ref{thm:operator_properties}.

We next establish a bound uniform in $h$. By
Assumption~\ref{as:regularity},
\[
m_\ell\le \ell(x)\le M_\ell.
\]
If $m_\ell\le v,w\le M_\ell$, then
\begin{equation}
m_\ell
\le
a_hw(x)+b_h\mathcal M_{\Delta t_{\mathrm{space}}}[v](x)
\le
M_\ell,
\end{equation}
since $a_h,b_h\ge0$ and $a_h+b_h=1$.
Thus $[m_\ell,M_\ell]$ is invariant under
$\mathcal T_h^w$. Starting from
$V_N^h=\ell$ and proceeding backward in time gives
\begin{equation}
m_\ell
\le
V_i^h(x)
\le
M_\ell
\qquad
\text{for all }x,\ i,\ h.
\label{eq:app_stability}
\end{equation}
Hence the family of exact discrete solutions is uniformly bounded.
Each $V_i^h$ is also bounded and uniformly continuous in $x$: the one-step state-transition maps are Lipschitz continuous in \(x\), uniformly in \(u\in\mathcal U\), with Lipschitz constant \(1+L_f\Delta t_{\mathrm{space}}\).,
so $\mathcal T_h^w$ preserves bounded uniform continuity whenever
$w$ has this property. Its iterates converge uniformly, and backward
induction starts from the Lipschitz function $\ell$. Thus the suprema
in the discrete solution are attained.

\paragraph{Step 2: Consistency with the terminal condition.}

Although \eqref{eq:app_terminal} imposes the terminal value exactly
on the grid, we additionally show that values approaching the terminal
time converge uniformly to $\ell$.

Set
\[
C_\ell:=L_\ell F_f,
\qquad
\underline V_i^h(x)
:=
\ell(x)-C_\ell(T-t_i).
\]
We claim that
\begin{equation}
\underline V_i^h(x)
\le
V_i^h(x)
\le
\ell(x).
\label{eq:app_barrier}
\end{equation}
The upper bound follows immediately from
\eqref{eq:app_scheme}. At $i=N$, the lower bound holds with
equality. Suppose now that
$\underline V_{i+1}^h\le V_{i+1}^h$. Using the Lipschitz continuity of $\ell$ and
$\|f(x,u)\|\le F_f$,
\begin{equation}
\ell(x+\Delta t_{\mathrm{space}} f(x,u))
\ge
\ell(x)-L_\ell F_f\Delta t_{\mathrm{space}},
\end{equation}
and hence
\begin{equation}
\mathcal M_{\Delta t_{\mathrm{space}}}[\underline V_i^h](x)
\ge
\underline V_i^h(x)-C_\ell\Delta t_{\mathrm{space}}.
\end{equation}
Moreover,
\[
\underline V_{i+1}^h(x)
=
\underline V_i^h(x)+C_\ell\Delta t_{\mathrm{time}}.
\]
Therefore,
\begin{equation}
\begin{aligned}
&a_hV_{i+1}^h(x)
+
b_h\mathcal M_{\Delta t_{\mathrm{space}}}[\underline V_i^h](x)
\\
&\quad\ge
a_h\underline V_{i+1}^h(x)
+
b_h\mathcal M_{\Delta t_{\mathrm{space}}}[\underline V_i^h](x)
\\
&\quad\ge
\underline V_i^h(x)
+
C_\ell
\left(
a_h\Delta t_{\mathrm{time}}-b_h\Delta t_{\mathrm{space}}
\right)
\\
&\quad=
\underline V_i^h(x).
\end{aligned}
\end{equation}
Since also $\underline V_i^h\le\ell$, we obtain
\[
\mathcal T_h^{V_{i+1}^h}[\underline V_i^h]
\ge
\underline V_i^h.
\]
By monotonicity of $\mathcal T_h^{V_{i+1}^h}$, iteration from
$\underline V_i^h$ produces a nondecreasing sequence converging to
its unique fixed point. Hence
$\underline V_i^h\le V_i^h$, proving
\eqref{eq:app_barrier}. Consequently,
\begin{equation}
0
\le
\ell(x)-V_i^h(x)
\le
L_\ell F_f(T-t_i).
\label{eq:app_terminal_consistency}
\end{equation}
Thus the discrete solutions converge uniformly to $\ell$ as
$t_i\nearrow T$. More precisely, for arbitrary discrete nodes with
$x_k\to x$ and $t_{i_k}\to T$,
\[
|V_{i_k}^{h_k}(x_k)-\ell(x)|
\le C_\ell(T-t_{i_k})+L_\ell\|x_k-x\|\longrightarrow0.
\]
Consequently, both half-relaxed limits attain the terminal data;
this also holds for the extension in \eqref{eq:app_time_extension}.

\paragraph{Step 3: Monotonicity in the sense of
Barles--Souganidis.}

To express \eqref{eq:app_scheme} in residual form, define
\begin{equation}
\begin{aligned}
\widehat S_h(x,t_i,r,\psi)
:=\max\Bigl\{&\lambda_h\bigl(r-\ell(x)\bigr),\\
&r-a_h\psi(x,t_i+\Delta t_{\mathrm{time}})\\
&\quad{}-b_h\sup_{u\in\mathcal U}
\psi\bigl(x+\Delta t_{\mathrm{space}}f(x,u),t_i\bigr)
\Bigr\}.
\end{aligned}
\label{eq:scheme_residual}
\end{equation}
The residual to which the specialized convergence theorem is applied is
\begin{equation}
S_h(x,t_i,r,\psi):=\lambda_h^{-1}\widehat S_h(x,t_i,r,\psi).
\label{eq:app_normalized_residual}
\end{equation}
Since $\lambda_h>0$, this normalization preserves the discrete equation
and the monotonicity inequalities.
Indeed, the fixed-point relation
\begin{equation}
\begin{aligned}
r=\min\Bigl\{\ell(x),\,
&a_h\psi(x,t_i+\Delta t_{\mathrm{time}})\\
&{}+b_h\sup_{u\in\mathcal U}
\psi\bigl(x+\Delta t_{\mathrm{space}}f(x,u),t_i\bigr)
\Bigr\}
\end{aligned}
\end{equation}
is equivalent to
\[
r-\ell(x)\le 0,
\]
and
\[
r
-
a_h\psi(x,t_i+\Delta t_{\mathrm{time}})
-
b_h\sup_{u\in\mathcal U}
\psi\bigl(x+\Delta t_{\mathrm{space}}f(x,u),t_i\bigr)
\le 0,
\]
with at least one of the two inequalities holding with equality. Since
$\lambda_h>0$, this is equivalent to
\[
\widehat S_h(x,t_i,r,\psi)=0,
\]
where $\widehat S_h$ is defined in~\eqref{eq:scheme_residual}.

Now let $\psi\ge\widetilde\psi$ pointwise. Then
\begin{equation}
\psi(x,t_i+\Delta t_{\mathrm{time}})
\ge
\widetilde\psi(x,t_i+\Delta t_{\mathrm{time}})
\end{equation}
and
\begin{equation}
\sup_u
\psi(x+\Delta t_{\mathrm{space}} f(x,u),t_i)
\ge
\sup_u
\widetilde\psi(x+\Delta t_{\mathrm{space}} f(x,u),t_i).
\end{equation}
It follows that
\begin{equation}
\widehat S_h(x,t_i,r,\psi)
\le
\widehat S_h(x,t_i,r,\widetilde\psi).
\label{eq:app_monotonicity}
\end{equation}
The same inequality holds for $S_h$. This is the monotonicity
convention used by \citet{barles1991convergence}.

\paragraph{Step 4: Consistency with the HJ variational inequality.}

We now verify consistency in the sense of
\citet{barles1991convergence}, using the main-text Hamiltonian
$H(p,x)=\max_{u\in\mathcal U}p^\top f(x,u)$.

Let
$\phi\in C_b^2(\mathbb R^n\times[0,T])$. Here bounded smooth test
functions suffice, since local tests can be extended outside a
neighborhood of the contact point without altering the calculation.
For $(x,t)$ in a compact subset of
$\mathbb R^n\times[0,T)$, Taylor expansion gives
\begin{equation}
\phi(x,t+\Delta t_{\mathrm{time}})
=
\phi(x,t)
+
\Delta t_{\mathrm{time}}\,D_t\phi(x,t)
+
O(\Delta t_{\mathrm{time}}^2).
\end{equation}
Similarly,
\begin{equation}
\phi(x+\Delta t_{\mathrm{space}} f(x,u),t)
=
\phi(x,t)
+
\Delta t_{\mathrm{space}}
\nabla\phi(x,t)^\top f(x,u)
+
O(\Delta t_{\mathrm{space}}^2).
\end{equation}
Because $\mathcal U$ is compact and $f$ is uniformly bounded,
the latter remainder is uniform in $u$. Therefore,
\begin{equation}
\begin{aligned}
&\max_{u\in\mathcal U}
\phi(x+\Delta t_{\mathrm{space}} f(x,u),t)
\\
&\qquad=
\phi(x,t)
+
\Delta t_{\mathrm{space}} H(\nabla\phi(x,t),x)
+
O(\Delta t_{\mathrm{space}}^2),
\end{aligned}
\end{equation}
where
\[
H(p,x)
:=
\max_{u\in\mathcal U}p^\top f(x,u).
\]

Using
$a_h\Delta t_{\mathrm{time}}=b_h\Delta t_{\mathrm{space}}=\lambda_h$,
we obtain
\begin{equation}
\begin{aligned}
&a_h\phi(x,t+\Delta t_{\mathrm{time}})
+
b_h
\max_u
\phi(x+\Delta t_{\mathrm{space}} f(x,u),t)
\\
&\qquad=
\phi(x,t)
+
\lambda_h
\left[
D_t\phi(x,t)
+
H(\nabla\phi(x,t),x)
\right]
\\
&\qquad\quad
+
O\!\left(
\lambda_h(\Delta t_{\mathrm{time}}+\Delta t_{\mathrm{space}})
\right).
\end{aligned}
\end{equation}

Consequently,
\begin{equation}
\begin{aligned}
&\frac{1}{\lambda_h}\Bigl[
\phi(x,t)-a_h\phi(x,t+\Delta t_{\mathrm{time}})\\
&\qquad\qquad{}
-b_h\max_u\phi(x+\Delta t_{\mathrm{space}}f(x,u),t)\Bigr]
\\
&\qquad=
-D_t\phi
-H(\nabla\phi,x)
+
O(\Delta t_{\mathrm{time}}+\Delta t_{\mathrm{space}}).
\end{aligned}
\label{eq:app_local_consistency}
\end{equation}

The expansion in (\ref{eq:app_local_consistency}) establishes consistency of the continuation branch when
the scheme is evaluated at a fixed smooth test function $\phi$ and at the same
point $(x,t)$. The Barles--Souganidis argument requires a stronger
form of consistency. In the viscosity argument, a smooth test function
$\phi$ touches a half-relaxed limit of the discrete solutions at $(x,t)$, while
the corresponding contact points for the discrete solutions generally occur at
nearby points $(x_k,t_k)$ rather than exactly at $(x,t)$. Moreover, the discrete
solution at the contact point may differ from the test function by a vanishing
constant offset. Accordingly, consider arbitrary sequences of nonterminal time-grid
nodes converging to $(x,t)$ with $0<t<T$:
\begin{equation}
h_k\to0,\qquad
(x_k,t_k)\to(x,t),\qquad
\xi_k\to0,
\end{equation}
and evaluate the scheme at
\begin{equation}
r=\phi(x_k,t_k)+\xi_k,
\qquad
\psi=\phi+\xi_k.
\end{equation}

For the continuation branch, the constant perturbation cancels exactly:
\begin{equation}
\begin{aligned}
&\phi(x_k,t_k)+\xi_k
-a_{h_k}\bigl(\phi(x_k,t_k+\Delta t_{\mathrm{time},k})+\xi_k\bigr) \\
&\qquad
-b_{h_k}\max_{u\in\mathcal U}
\bigl(
\phi(x_k+\Delta t_{\mathrm{space},k}f(x_k,u),t_k)+\xi_k
\bigr) \\
&=
\phi(x_k,t_k)
-a_{h_k}\phi(x_k,t_k+\Delta t_{\mathrm{time},k}) \\
&\qquad
-b_{h_k}\max_{u\in\mathcal U}
\phi(x_k+\Delta t_{\mathrm{space},k}f(x_k,u),t_k),
\end{aligned}
\end{equation}
because $a_{h_k}+b_{h_k}=1$. Hence the vanishing offset $\xi_k$ does not affect
the continuation residual, and the expansion in (\ref{eq:app_local_consistency}) remains valid along the
sequence $(x_k,t_k)$.

For the obstacle branch, continuity of $\phi$ and $\ell$, together with
$(x_k,t_k)\to(x,t)$ and $\xi_k\to0$, gives
\begin{equation}
\phi(x_k,t_k)+\xi_k-\ell(x_k)
\longrightarrow
\phi(x,t)-\ell(x).
\end{equation}
Combining the two branches gives
\begin{equation}
\begin{aligned}
&\frac{1}{\lambda_{h_k}}\widehat S_{h_k}
\bigl(x_k,t_k,\phi(x_k,t_k)+\xi_k,\phi+\xi_k\bigr)\\
&\qquad\longrightarrow
\max\Bigl\{\phi(x,t)-\ell(x),\,
-D_t\phi(x,t)-H\bigl(\nabla\phi(x,t),x\bigr)\Bigr\}.
\end{aligned}
\label{eq:bs_consistency_limit}
\end{equation}

This verifies consistency for the normalized residual $S_h$:
its value on smooth test functions converges to the limiting residual
along every such sequence, including arbitrary vanishing constant
perturbations \citep{barles1991convergence}. Therefore, the limiting
equation associated with the proposed scheme is
\begin{equation}
\max\left\{
V-\ell,
-D_tV-H(\nabla V,x)
\right\}
=0.
\label{eq:app_max_HJ}
\end{equation}
We use the standard viscosity convention for
\eqref{eq:app_max_HJ}: its residual is nonpositive at upper test
contacts and nonnegative at lower test contacts. Since
$-\max\{a,b\}=\min\{-a,-b\}$, its minimum-form expression is
\begin{equation}
\min\left\{
\ell-V,
D_tV+H(\nabla V,x)
\right\}
=0,
\label{eq:app_HJVI}
\end{equation}
which has the same expression as \eqref{eq:HJR}. To preserve the
same viscosity solution, the residual in this minimum-form terminal
equation is interpreted as nonnegative at upper test contacts and
nonpositive at lower test contacts. Thus the two forms are equivalent
with these corresponding conventions; the convergence theorem is
applied to the maximum form throughout.

\paragraph{Step 5: Comparison and identification of the limit.}

Under Assumption~\ref{as:regularity}, the Hamiltonian is continuous.
Indeed,
\begin{equation}
|H(p,x)-H(q,x)|
\le
F_f\|p-q\|,
\label{eq:app_H_gradient_bound}
\end{equation}
and
\begin{equation}
|H(p,x)-H(p,y)|
\le
L_f\|p\|\,\|x-y\|.
\label{eq:app_H_state_bound}
\end{equation}
Moreover, $\ell$ is bounded and uniformly continuous.
We invoke the comparison principle of
\cite[Theorem~2]{hirsch2026extending}.
Define $\mathcal H(x,t,p):=H(p,x)$.
The two Hamiltonian estimates above imply conditions~(i)--(ii)
of that theorem, since
$1+\lVert x\rVert\ge 1$ and
$1+\lVert p\rVert\ge\lVert p\rVert$.

In the notation of \cite{hirsch2026extending}, take
$r\equiv-\infty$, $q(x,t)=\ell(x)$, and terminal function $\ell$.
The choice $r\equiv-\infty$ is permitted by their
Definition~2. With these choices, their GRA-TVP reduces to
\[
-\min\bigl\{
\ell(x)-V,\,
\partial_t V+H(\nabla V,x)
\bigr\}=0,
\qquad
V(\cdot,T)=\ell,
\]
which is exactly the maximum-form problem established above.
Therefore, taking $S=0$ in their Theorem~2 gives comparison
on $\mathbb R^n\times(0,T]$ between upper-semicontinuous
subsolutions and lower-semicontinuous supersolutions
satisfying the terminal inequalities.

By the preceding steps, the upper and lower half-relaxed
limits $\overline V$ and $\underline V$ are bounded and are,
respectively, a viscosity subsolution and a viscosity
supersolution of this problem. The terminal estimate already
proved gives
\[
\overline V(x,T)=\underline V(x,T)=\ell(x).
\]
Their extensions by $\ell$ at $t=T$ retain the respective
semicontinuity and satisfy the terminal conditions in
\cite[Definition~5]{hirsch2026extending}.
Comparison consequently yields
$\overline V\le\underline V$ on
$\mathbb R^n\times(0,T]$.
Since $\underline V\le\overline V$ by definition, the two
limits coincide. Their common value is continuous and is
the unique bounded continuous viscosity solution.
This provides the semicontinuous comparison required by
the monotone approximation theorem.

We have now verified the hypotheses of the monotone approximation
theorem recalled above:

\begin{enumerate}
    \item the exact discrete solution exists and is unique at each
          time slice by contraction;
    \item the scheme is uniformly stable by
          \eqref{eq:app_stability};
    \item the normalized residual is monotone by
          \eqref{eq:app_normalized_residual}--\eqref{eq:app_monotonicity};
    \item consistency holds for the maximum-form HJ variational
          inequality by \eqref{eq:bs_consistency_limit};
    \item the terminal traces are recovered by
          \eqref{eq:app_terminal_consistency};
    \item the limiting problem satisfies comparison in the bounded
          upper-/lower-semicontinuous classes specified above.
\end{enumerate}

Therefore, for every sequence
\[
h_k=(\Delta t_{\mathrm{time},k},\Delta t_{\mathrm{space},k})\to(0,0),
\]
Theorem~\ref{thm:barles_souganidis_specialized}, applied to $S_h$,
gives local uniform convergence to the unique continuous viscosity
solution $V$ of \eqref{eq:app_max_HJ} on
$\mathbb R^n\times(0,T]$.

The endpoint $t=0$ follows from the same contraction estimate.
For $i\le N-2$, comparing the fixed-point equations at two adjacent
slices and using $1-b_h=a_h$ gives
\[
\|V_i^h-V_{i+1}^h\|_\infty
\le \|V_{i+1}^h-V_{i+2}^h\|_\infty.
\]
Together with \eqref{eq:app_terminal_consistency} at $i=N-1$, this
bounds every adjacent-slice difference by
$C_\ell\Delta t_{\mathrm{time}}$. Consequently,
\[
\|V^h(\cdot,t)-V^h(\cdot,s)\|_\infty
\le C_\ell\bigl(|t-s|+\Delta t_{\mathrm{time}}\bigr).
\]
Passing to the limit for positive times and then letting the times
approach zero extends both $V$ continuously and the convergence locally
uniformly to $t=0$. Since the limit is unique, the entire family satisfies
\begin{equation}
V^h
\longrightarrow
V
\qquad
\text{locally uniformly on }\mathbb R^n\times[0,T]
\text{ as }h\to(0,0).
\label{eq:app_final_convergence}
\end{equation}

Finally, by the standard viscosity characterization of finite-horizon
HJ reachability
\citep{lygeros2004reachability,mitchell2005time},
this limit is the finite-horizon HJR value function. Equivalently,
it solves \eqref{eq:HJR} under the terminal viscosity convention
specified after \eqref{eq:app_HJVI}. Hence the exact discrete
fixed-point solution converges to the original HJR value function.
\end{proof}
\begin{remark}[Scope of the convergence result]
Theorem~\ref{thm:convergence_to_viscosity} concerns the exact fixed
points of the discrete operator. Neural function approximation,
adaptive zero-level-set sampling, target-network updates, and the
additional conservative fusion
\[
V_i
=
\min\left\{
\widehat V_i,
\widehat V_i^{\mathrm{DP}}
\right\}
\]
are implementation-level approximations and are not required for the
consistency result above. Their contribution to the total error is
treated separately as neural approximation error.
\end{remark}

\begin{remark}[Discretization ratio]
The convergence result in Theorem~3 does not require
$\Delta t_{\mathrm{time}}/\Delta t_{\mathrm{space}}$ to remain fixed.
It is sufficient that
$\Delta t_{\mathrm{time}}\to0$ and
$\Delta t_{\mathrm{space}}\to0$, since the normalized truncation error is
$O(\Delta t_{\mathrm{time}}+\Delta t_{\mathrm{space}})$.
The ratio between the two discretization steps only affects the contraction
modulus
\[
b_h=
\frac{\Delta t_{\mathrm{time}}}
{\Delta t_{\mathrm{time}}+\Delta t_{\mathrm{space}}},
\]
and hence the convergence rate of the inner fixed-point iteration.
\end{remark}
\newpage
\section{Experiment Details}\label{app:exp-details}
\subsection{Contract2Reach Implementation Details}

Algorithm~\ref{alg:contract2reach} gives the full procedure and
Table~\ref{tab:hyperparams} lists its hyperparameters. Every symbol introduced below
appears in both.

\begin{algorithm2e}[H]
\SetAlgoVlined
\LinesNumbered
\SetCommentSty{emph}
\caption{Contract2Reach}
\label{alg:contract2reach}
\KwIn{horizon $T$, space step $\Delta t_{\mathrm{space}}$, space steps per time step $m$, iterations per slice $E$, Polyak coefficient $\kappa$; remaining hyperparameters in Table~\ref{tab:hyperparams}}
\KwOut{value slices $V_{\theta_i}\approx V(\cdot,t_i)$ at $t_i=i\,\Delta t_{\mathrm{time}}$, $i=0,\dots,N$}
\BlankLine
$\Delta t_{\mathrm{time}}\leftarrow m\,\Delta t_{\mathrm{space}}$, \ $N\leftarrow T/\Delta t_{\mathrm{time}}$, \ $\beta\leftarrow m/(m+1)$\nllabel{ln:init} \tcp*[r]{contraction factor}
\tcp{Terminal slice at $t_N=T$: regress the constraint}
train $V_{\theta_N}$ to fit $\ell$ on states drawn from $\mathcal S_0$ and near $\{\ell=0\}$, then freeze $\theta_N$\nllabel{ln:pool0}\;
\For(\tcp*[f]{one slice at a time, backward in time}){$i=N-1,N-2,\dots,0$\nllabel{ln:outer}}{
  $\theta_i\leftarrow\theta_{i+1}$, \ $\bar\theta_i\leftarrow\theta_i$, \ clear the replay buffer\nllabel{ln:warm} \tcp*[r]{warm start}
  \For{$e=1,\dots,E$}{
    \If{$e \bmod M_c=0$\nllabel{ln:collect}}{
      roll out $T_r=m\,\min(n_{\mathrm{win}},N-i)$ space steps from initial states in the pool and uniform samples, choosing each control from the slice at that time: $V_{\theta_i}$ for the first $m$ steps and frozen $V_{\theta_{i+1}},V_{\theta_{i+2}},\dots$ afterwards\nllabel{ln:rollout}\;
      compute $\hat V$ with the target network $V_{\bar\theta_i}$ (Eq.~\eqref{eq:vhat}) and $\hat V^{\mathrm{DP}}$ from frozen slices only (Eq.~\eqref{eq:vdp})\nllabel{ln:targets}\;
      store each $x_0$ with its target $y$ from $\mathcal C(\hat V,\hat V^{\mathrm{DP}})$ and $\ell(x_0)$ (Eq.~\eqref{eq:term})\nllabel{ln:y}\;
    }
    take one Adam step on $\mathrm{LSE}_\lambda\bigl((V_{\theta_i}(x_0)-y)^2\bigr)$ over a minibatch\nllabel{ln:minibatch} \tcp*[r]{single regression}
    $\bar\theta_i\leftarrow(1-\kappa)\,\bar\theta_i+\kappa\,\theta_i$\nllabel{ln:update} \tcp*[r]{Polyak averaging}
    \If{$e \bmod M_p=0$}{
      refresh the pool with bisection points on $\{V_{\bar\theta_i}=0\}$ (Section~\ref{sec:sampling})\nllabel{ln:refresh}\;
    }
  }
  freeze $\theta_i$ \tcp*[r]{slice $i$ solved; used to train slice $i-1$}
}
\Return{$\{\theta_i\}_{i=0}^{N}$}
\end{algorithm2e}

\paragraph{Problem form.}
We use time-to-go coordinates $\tau=T-t$, so the terminal
condition is imposed at $\tau=0$.
The value functions are
\begin{equation}
V(x,\tau)=
\begin{cases}
\displaystyle
\sup_{u(\cdot)}\min_{s\in[0,\tau]}
\ell\bigl(\xi_x^u(s)\bigr),
& \text{PVD 2D, Quadrotor 13D},\\[5pt]
\displaystyle
\sup_{u(\cdot)}\max_{s\in[0,\tau]}
\ell\bigl(\xi_x^u(s)\bigr),
& \text{PubSub 40D}.
\end{cases}
\end{equation}
The extrema over controls are taken over admissible control signals.
The minimum over time enforces safety throughout the horizon,
whereas the maximum permits reaching the target at any time.
The control maximizes the value in all three problems.
For the exact value function, $\{x:V(x,\tau)>0\}$ is the
finite-horizon safe set in the avoidance problems and the
reachable set in the PubSub problem.


\begin{table}[!t]
\centering
\caption{Hyperparameters and sign conventions of Contract2Reach.
The notation follows Algorithm~\ref{alg:contract2reach}.}
\label{tab:hyperparams}
\footnotesize
\setlength{\tabcolsep}{4pt}
\resizebox{\linewidth}{!}{%
\begin{tabular}{llccc}
\toprule
Quantity & Symbol & PVD 2D & Quadrotor 13D & PubSub 40D \\
\midrule

\multicolumn{5}{l}{\emph{Problem}} \\
Problem type
& --
& avoid & avoid & reach \\

Meaning of $\ell>0$
& --
& safe & safe & target \\

Value function
& $V$
& $\sup_u \min_{s\in[0,\tau]} \ell$
& $\sup_u \min_{s\in[0,\tau]} \ell$
& $\sup_u \max_{s\in[0,\tau]} \ell$ \\

Time horizon
& $T$
& 1.2 s & 1.0 s & 1.0 s \\

\midrule
\multicolumn{5}{l}{\emph{Discretization}} \\

Space step (dynamics integration)
& $\Delta t_{\mathrm{space}}$
& 0.005 s & 0.005 s & 0.005 s \\

Space steps per time step
& $m$
& 4 & 4 & 4 \\

Time step (between slices)
& $\Delta t_{\mathrm{time}}=m\,\Delta t_{\mathrm{space}}$
& 0.02 s & 0.02 s & 0.02 s \\

Number of slices
& $K+1=T/\Delta t_{\mathrm{time}}+1$
& 61 & 51 & 51 \\

Contraction factor
& $\beta=m/(m+1)$
& 0.8 & 0.8 & 0.8 \\

\midrule
\multicolumn{5}{l}{\emph{Target construction}} \\

Lookahead (space steps)
& $L$
& 4 & 4 & 4 \\

Anchor window (time steps)
& $n_{\mathrm{win}}$
& 10 & 20 & 10 \\

Maximum rollout length (space steps)
& $m\,n_{\mathrm{win}}$
& 40 & 80 & 40 \\

Target combination
& $\mathcal C$
& $\hat V$
& $\min(\hat V,\hat V^{\mathrm{DP}})$
& $\min(\hat V,\hat V^{\mathrm{DP}})$ \\

\midrule
\multicolumn{5}{l}{\emph{Value network and optimization}} \\

Hidden layers (tanh)
& --
& $3\times256$ & $3\times512$ & $3\times512$ \\

Ensemble size
& --
& 2 & 2 & 2 \\

Learning rate (Adam)
& $\eta$
& $3\times10^{-4}$ & $3\times10^{-4}$ & $3\times10^{-4}$ \\

Polyak coefficient
& $\kappa$
& 0.005 & 0.005 & 0.005 \\

Soft-$L_\infty$ temperature
& $\lambda$
& 20 & 50 & 50 \\

Iterations per slice
& $E$
& 30{,}000 & 30{,}000 & 30{,}000 \\

\midrule
\multicolumn{5}{l}{\emph{Data collection and sampling}} \\

Rollouts per collection
& $N_c$
& 32{,}768 & 65{,}536 & 65{,}536 \\

Minibatch size
& $B$
& 32{,}768 & 131{,}072 & 131{,}072 \\

Replay buffer (collections)
& $D$
& 4 & 4 & 4 \\

Collection period (iters)
& $M_c$
& 500 & 250 & 250 \\

Pool size
& $N_p$
& 32{,}768 & 65{,}536 & 65{,}536 \\

Pool share of initial states
& $\rho$
& 0.7 & 0.7 & 0.7 \\

Pool refresh period (iters)
& $M_p$
& 500 & 500 & 500 \\

Near-boundary share, first slice
& $\alpha_0$
& 0 & 0.75 & 0.75 \\

Complement source, first slice
& $\mathcal S_0$
& uniform & uniform & radial \\

Near-boundary share, later slices
& $\alpha$
& 0 & 0.6 & 0.6 \\

Complement source, later slices
& $\mathcal S$
& visited & visited & visited \\

Input normalization
& --
& none & box to $[-1,1]$ & metric-aligned scaling \\

\bottomrule
\end{tabular}%
}
\end{table}

\newpage

\paragraph{Two time grids.}
The scheme uses two step sizes. The \emph{space step} $\Delta t_{\mathrm{space}}$ is the integration step of the dynamics. Rollouts advance one space step at a time, and we index these steps by $j$. The \emph{time step} $\Delta t_{\mathrm{time}}$ separates consecutive value slices. We index slices by $i$, so that slice $i$ lies at $t_i=i\,\Delta t_{\mathrm{time}}$ and the terminal slice lies at $t_N=T$. The two step sizes are related by $\Delta t_{\mathrm{time}}=m\,\Delta t_{\mathrm{space}}$ with $m=4$. One time step therefore contains four space steps, and space step $j$ of a rollout started at slice $i$ lies at slice $i+\lfloor j/m\rfloor$. The ratio of the two step sizes fixes the contraction factor $\beta=\Delta t_{\mathrm{time}}/(\Delta t_{\mathrm{time}}+\Delta t_{\mathrm{space}})=m/(m+1)=0.8$ (line~\ref{ln:init}), so $\beta$ is set by the discretization rather than tuned.

\paragraph{Slice-by-slice training.}
Slices are computed backward in time, in the order $i=N-1,\dots,0$ (line~\ref{ln:outer}). The terminal slice $i=N$ regresses the terminal condition $V(\cdot,T)=\ell$ (line~\ref{ln:pool0}). Slice $i$ is initialized from slice $i+1$ (line~\ref{ln:warm}), trained, and then frozen. Slices solved after it only read it. Every slice before the terminal one is trained on targets built from rollouts, as described next.

\paragraph{Rollouts.}
All three benchmarks maximize the value over admissible controls. For the control-affine dynamics $f(x,u)=f_0(x)+G(x)u$ with componentwise bounds $u_{\min}\le u\le u_{\max}$, the rollout control is chosen greedily with respect to the learned value. No policy network is trained.

A rollout for slice $i$ runs $T_r=m\,n$ integration steps, where $n=\min(n_{\mathrm{win}},N-i)$. At step $j=0,\ldots,T_r-1$, let $s_j=i+\lfloor j/m\rfloor$ and choose
\[
u_j
=
u_{\mathrm{mid}}
+
u_{\mathrm{rad}}\odot
\mathrm{sign}\!\left(
G(x_j)^\top\nabla V_{\theta_{s_j}}(x_j)
\right),
\]
where
\[
u_{\mathrm{mid}}=\frac{u_{\max}+u_{\min}}{2},
\qquad
u_{\mathrm{rad}}=\frac{u_{\max}-u_{\min}}{2}.
\]
This maximizes the Hamiltonian associated with the selected value estimate. For symmetric bounds, it reduces to $\mathrm{sign}(G^\top\nabla V_{\theta_{s_j}})\odot u_{\max}$.

The current slice $V_{\theta_i}$ is used for the first $m$ steps, and the frozen slices $V_{\theta_{i+1}},V_{\theta_{i+2}},\dots$ are used thereafter. Each control is therefore computed from the slice at the time the rollout has reached (line~\ref{ln:rollout}).
 
\paragraph{Targets.}
Each rollout $x_0,\dots,x_{T_r}$ yields two targets for $V_{\theta_i}(x_0)$ (line~\ref{ln:targets}). The \emph{dynamic-programming target} follows the rollout to its end and reads the value there from the slice $n$ time steps ahead,
\begin{equation}
\hat V^{\mathrm{DP}}(x_0)=
\begin{cases}
\displaystyle
\min\left\{
\min_{0\le j\le T_r}\ell(x_j),\,
V_{\theta_{i+n}}(x_{T_r})
\right\},
& \text{PVD 2D, Quadrotor 13D},\\[5pt]
\displaystyle
\max\left\{
\max_{0\le j\le T_r}\ell(x_j),\,
V_{\theta_{i+n}}(x_{T_r})
\right\},
& \text{PubSub 40D},
\end{cases}
\label{eq:vdp}
\end{equation}
where $V_{\theta_{i+n}}$ is replaced by $\ell$ when $i+n=N$. It never queries the network being trained. The \emph{contractive target} looks $L$ space steps ahead and closes the recursion with $V_{\bar\theta}$, a Polyak-averaged copy of the network being trained:
Set
\[
v_L=V_{\bar\theta_i}(x_L),
\qquad
\hat V(x_0)=v_0.
\]
For $j=L-1,\ldots,0$, define
\begin{equation}
v_j=
\begin{cases}
\min\left\{
\ell(x_j),(1-\beta)R_j+\beta v_{j+1}
\right\},
& \text{PVD 2D, Quadrotor 13D},\\[3pt]
\max\left\{
\ell(x_j),(1-\beta)R_j+\beta v_{j+1}
\right\},
& \text{PubSub 40D}.
\end{cases}
\label{eq:vhat}
\end{equation}
where $R_j$ is the right-hand side of~\eqref{eq:vdp} evaluated from $x_j$ over a window one time step shorter. The network's own prediction enters $\hat V$ with weight $\beta^{L}$. With $L=4$ this weight is $0.41$.

\paragraph{Combining the targets.}
In the Quadrotor 13D and the PubSub 40D, the control maximizes the value.
The rollout follows one admissible control and therefore attains at most the optimal value, so $\hat V^{\mathrm{DP}}$ is a lower bound. We use
$\mathcal{C}=\min(\hat V,\hat V^{\mathrm{DP}})$, which lowers the target and shrinks the certified set $\{V>0\}$, the conservative direction. For the PVD 2D, we use $\hat V$ alone. Finally, since $s=0$ is included in the maximum or minimum over time, the regression target is combined with $\ell(x_0)$ (line~\ref{ln:y}),
\begin{equation}
y=
\begin{cases}
\min\left\{\mathcal C,\ell(x_0)\right\},
& \text{PVD 2D, Quadrotor 13D},\\[3pt]
\max\left\{\mathcal C,\ell(x_0)\right\},
& \text{PubSub 40D}.
\end{cases}
\label{eq:term}
\end{equation}
 
\paragraph{Optimization.}
Each slice is trained for $E$ iterations of Adam with learning rate $\eta$, starting from the weights of the previous slice and a fresh optimizer state, and the averaged copy is updated as $\bar\theta\leftarrow(1-\kappa)\bar\theta+\kappa\theta_i$ (line~\ref{ln:update}). The value network is an ensemble of two $\tanh$ multilayer perceptrons. Errors matter most near the zero level set, where they are also largest, so instead of the mean squared
error we minimize its soft maximum over a minibatch of size $B$,
\begin{equation}
  \mathrm{LSE}_\lambda(e^2)=\frac{1}{\lambda}\log\Big(\frac{1}{B}\sum_{k=1}^{B}
  \exp\big(\lambda\, e_k^2\big)\Big),
  \label{eq:lse}
\end{equation}
which reduces to the mean squared error as $\lambda\to0$ and to the largest squared error
as $\lambda\to\infty$.

\paragraph{Data collection.}
Training data for slice $i$ are produced in \emph{collections}
(lines~\ref{ln:collect}--\ref{ln:targets}). A collection draws $N_c$ initial states: a share $\rho$ comes from an initial-state pool $\mathcal{P}$, described below, and the remaining $(1-\rho)N_c$ are sampled uniformly over the domain. Each initial state is
rolled out for $T_r$ space steps with the time-varying policy, and the two targets $\hat V^{\mathrm{DP}}(x_0)$ and $\hat V(x_0)$ are computed from its rollout. Only the initial states and their targets are stored; the intermediate rollout states are used to compute the targets and to refresh the pool, but are not regressed on directly. The stored
pairs enter a replay buffer $\mathcal{R}$ that keeps the latest $D$ collections, i.e., at most $D\,N_c$ pairs. The buffer is emptied at the start of each slice and filled by collecting at every iteration until it holds $D$ collections; afterwards a new collection replaces the oldest one every $M_c$ iterations. Training targets are computed using the networks available at data collection.
 
\paragraph{Minibatch sampling.}
At every iteration we draw $B$ pairs uniformly from $\mathcal{R}$ and take one gradient step on them (lines~\ref{ln:minibatch}--\ref{ln:update}). We keep $B<D\,N_c$, a quarter of the buffer for the PVD 2D and half of it for the other two systems, so that no step fits
the entire buffer; in our experiments, fitting all stored targets at every step led the network to memorize stale targets.
 
\paragraph{Initial-state pool.}
The pool $\mathcal{P}$ holds $N_p$ states and steers a share $\rho$ of every collection toward the boundary of the certified set, where accuracy matters most, while the uniform share $1-\rho$ keeps the value function constrained elsewhere. Each collection draws $\rho N_c$ distinct pool states, so $N_p\ge\rho N_c$. The pool mixes near-boundary
states with states from a \emph{complement source}, the distribution that supplies the remaining pool entries. Its composition differs between the first slice and later slices, because the boundary is known in closed form only at $i=N$.
 
\emph{First slice ($i=N$).} Here $V=\ell$, so the boundary is the zero level set of $\ell$ and can be sampled directly (line~\ref{ln:pool0}). A share $\alpha_0$ of the pool is drawn in a thin band around $\{\ell=0\}$. The remaining share $1-\alpha_0$ comes from the complement source $\mathcal{S}_0$. For the PVD 2D and the Quadrotor 13D, $\mathcal{S}_0$ is the uniform distribution over the domain. For the PubSub 40D, it is a radial distribution: $\ell$ depends on the state through $q=x_0^2+\frac{1}{39}\sum_{i=1}^{39}x_i^2$, and we draw states whose value of $q$
is uniform on an interval $[0,q_{\max}]$. Uniform sampling of the $40$-dimensional box would instead place almost all states at large $q$, far from the target.
 
\emph{Later slices ($i<N$).} The boundary is now the zero level set of the network being trained and has to be located numerically. Every $M_p$ iterations the pool is rebuilt (line~\ref{ln:refresh}). States visited by recent rollouts are split by the sign of $V_{\bar\theta_i}$, segments joining states of opposite sign are bisected until they locate
the sign change, and the resulting points on $\{V_{\bar\theta_i}=0\}$ form a share $\alpha$ of the pool. The complement source $\mathcal{S}$ consists of \emph{visited states}, the states along recent rollouts taken every $m$ space steps. These are the states that the current policy actually reaches and at which the targets query the frozen slices. With
$\rho=0.7$ and $\alpha=0.6$, the initial states of a collection are therefore $42\%$ boundary points, $28\%$ visited states, and $30\%$ uniform states.

\subsection{Parameterized 2D Drone Details}
\paragraph{System dynamics.}
We consider the following parameterized 2D drone dynamics for longitudinal motion along the $z$-axis:
\begin{equation}
\dot x
=
f(x,u)
=
\begin{bmatrix}
v_z\\
Ku-g\\
0
\end{bmatrix},
\qquad
x=
\begin{bmatrix}
z\\
v_z\\
K
\end{bmatrix},
\qquad
u\in[-1,1].
\end{equation}
where $z$ denotes the vertical position, $v_z$ is the vertical velocity,
$g=9.8m/s^2$ is the gravitational acceleration, $K$ is a constant control-gain
parameter, and $u\in[-1,1] m/s^2$ is the control input. We implement the continuous-time
dynamics in sampled data with $\Delta t_{space}=0.005$ sec.
\paragraph{Safe set.}
The failure set is defined as
\[
\mathcal L
=
\left\{
x:\lvert z-1.5\rvert\ge1.5
\right\},
\]
corresponding to collision with the ground at $z=0$ or the ceiling
at $z=3$. The instantaneous safe region is therefore
\[
\mathcal S=\{x:0<z<3\}.
\]
Consistent with the avoidance convention in the main text, define
\[
\ell(x)=\min\{z,\,3-z\}=1.5-\lvert z-1.5\rvert.
\]
Then
\[
\mathcal L=\{x:\ell(x)\le0\},
\qquad
\mathcal S=\{x:\ell(x)>0\}.
\]
Equivalently, the two safety constraints are
\[
\ell_1(x)=z,
\qquad
\ell_2(x)=3-z,
\]
and safety requires both constraints to be strictly positive.
The specification directly constrains only the altitude $z$;
the velocity $v_z$ and parameter $K$ affect whether safety can
be maintained over the finite horizon.
\subsection{13D Quadrotor Details}
\paragraph{System dynamics.}
We consider the following 13-dimensional quadrotor dynamics:
\[
x=
\begin{bmatrix}
p_x & p_y & p_z &
q_w & q_x & q_y & q_z &
v_x & v_y & v_z & \omega_x & \omega_y & \omega_z
\end{bmatrix}^{\top},
\]
with control input
\[
u=
\begin{bmatrix}
F & \alpha_x & \alpha_y & \alpha_z
\end{bmatrix}^{\top}.
\]
The continuous-time dynamics are given by
\begin{equation}
\begin{aligned}
\dot p_x &= v_x,\quad
\dot p_y = v_y,\quad
\dot p_z = v_z,\\
\dot q_w &=
-\frac{1}{2}
\left(
\omega_x q_x+\omega_y q_y+\omega_z q_z
\right),\\
\dot q_x &=
\frac{1}{2}
\left(
\omega_x q_w+\omega_z q_y-\omega_y q_z
\right),\\
\dot q_y &=
\frac{1}{2}
\left(
\omega_y q_w-\omega_z q_x+\omega_x q_z
\right),\\
\dot q_z &=
\frac{1}{2}
\left(
\omega_z q_w+\omega_y q_x-\omega_x q_y
\right),\\
\dot v_x &=
\frac{C_T F}{m}
\left(
2q_wq_y+2q_xq_z
\right),\\
\dot v_y &=
\frac{C_T F}{m}
\left(
-2q_wq_x+2q_yq_z
\right),\\
\dot v_z &=
G_z-
\frac{C_T F}{m}
\left(
2q_x^2+2q_y^2-1
\right),\\
\dot\omega_x &= \alpha_x-\frac{5}{9}\omega_y\omega_z,\\
\dot\omega_y &= \alpha_y+\frac{5}{9}\omega_x\omega_z,\\
\dot\omega_z &= \alpha_z.
\end{aligned}
\label{eq:quadrotor_dynamics}
\end{equation}
Here $(p_x,p_y,p_z)$ and $(v_x,v_y,v_z)$ denote the position and
linear velocity, respectively,
$q=(q_w,q_x,q_y,q_z)$ is the unit quaternion representing the attitude,
and $(\omega_x,\omega_y,\omega_z)$ denotes the angular velocity.
The input $F$ is the collective thrust, while
$(\alpha_x,\alpha_y,\alpha_z)$ are the commanded angular accelerations.
We use $C_T=1$ and $m=1\,\mathrm{kg}$. We implement the continuous-time
dynamics in sampled data with $\Delta t_{space}=0.005$ sec.

\paragraph{Safe set.}
The safety specification requires the quadrotor to avoid a cylindrical
obstacle of radius $r_o=0.5\,\mathrm{m}$ centered at the origin and extending
along the $z$-axis. To account for the finite size and attitude of the
quadrotor, we first compute its body $z$-axis expressed in the inertial frame,
\[
\nu
=
\begin{bmatrix}
\nu_x & \nu_y & \nu_z
\end{bmatrix}^{\top}
=
q\,e_3\,\bar q,
\qquad
e_3=
\begin{bmatrix}
0 & 0 & 1
\end{bmatrix}^{\top},
\]
where $\bar q$ denotes the quaternion conjugate. The attitude-dependent
projected dimensions are
\[
d_x
=
r_a^2
\frac{p_x^2\nu_z^2}
{p_x^2\nu_x^2+p_x^2\nu_z^2
+2p_xp_y\nu_x\nu_y
+p_y^2\nu_y^2+p_y^2\nu_z^2},
\]
and
\[
d_y
=
r_a^2
\frac{p_y^2\nu_z^2}
{p_x^2\nu_x^2+p_x^2\nu_z^2
+2p_xp_y\nu_x\nu_y
+p_y^2\nu_y^2+p_y^2\nu_z^2},
\]
where $r_a=0.17\,\mathrm{m}$ is the radius of the quadrotor.

The safety function is then defined as
\begin{equation}
\ell(x)
=
\max\left\{
\sqrt{p_x^2+p_y^2}-\sqrt{d_x+d_y},
\,0
\right\}
-r_o.
\label{eq:quadrotor_safety}
\end{equation}
Accordingly, the failure set and safe region are
\[
\mathcal L
=
\left\{
x:\ell(x)\le0
\right\},
\qquad
\mathcal S
=
\left\{
x:\ell(x)>0
\right\}.
\]
Equivalently, we define the safety constraint
\[
h_S(x)=\ell(x),
\]
so that the interior safe region is characterized by $h_S(x)>0$.
The safety specification therefore captures collision with the cylindrical
obstacle while accounting for the finite radius and attitude of the quadrotor.
\subsection{40D Publisher-Subscriber Details}
\label{app:pubsub}
\paragraph{System dynamics.}
We consider the following 40-dimensional publisher--subscriber system:
\[
x=
\begin{bmatrix}
x_0 & x_1 & \cdots & x_{39}
\end{bmatrix}^{\top}
\in[-1,1]^{40},
\]
with control input
\[
u=
\begin{bmatrix}
u_1 & \cdots & u_{39}
\end{bmatrix}^{\top},
\qquad
u_i\in[-0.5,0.5].
\]
The dynamics are
\begin{equation}
\begin{aligned}
\dot x_0
&=a x_0,\\
\dot x_i
&=-x_0+a x_i+b u_i-\gamma x_0^2x_i,
\qquad i=1,\ldots,39,
\end{aligned}
\end{equation}
where
\[
a=-0.5,\qquad
b=0.4,\qquad
\gamma=20.
\]
Thus, $x_0$ represents the publisher state and each $x_i$,
$i=1,\ldots,39$, represents a subscriber state controlled independently
through $u_i$. We implement the continuous-time dynamics in sampled data
with $\Delta t_{\mathrm{space}}=0.005\,\mathrm{s}$.

\paragraph{Target set.}
Unlike the previous two avoidance problems, the publisher--subscriber
benchmark is formulated as a reachability problem. The target function is
defined as
\begin{equation}
\ell(x)
=
\frac{1}{2}
\left(
r^2-x_0^2-\frac{1}{39}\sum_{i=1}^{39}x_i^2
\right),
\qquad
r=0.25.
\end{equation}
Accordingly, the target set is
\[
\mathcal T
=
\left\{
x:
x_0^2+\frac{1}{39}\sum_{i=1}^{39}x_i^2<r^2
\right\}
=
\left\{
x:\ell(x)>0
\right\}.
\]
The zero level set $\ell(x)=0$ therefore defines the boundary of the
40-dimensional ellipsoidal target region. The reachability objective is to
determine the set of initial states from which the system can enter
$\mathcal T$ within the prescribed finite horizon.

\paragraph{Two-dimensional visualization of the 40D value function.}
The publisher--subscriber benchmark has a 40-dimensional state, and hence
the learned value function cannot be visualized directly over its full state
space. We therefore visualize two-dimensional restrictions of the
40-dimensional value function rather than constructing a reduced-order
two-dimensional dynamical system.

Specifically, let $m\in\{1,\ldots,39\}$ denote the number of subscriber
states that are varied together, and let $c\in[-1,1]$ denote the fixed value
assigned to the remaining subscribers. For visualization coordinates
$(X,Y)\in[-1,1]^2$, we construct the full 40-dimensional state
\begin{equation}
x^{m,c}(X,Y)
=
\left[
X,\,
\underbrace{Y,\ldots,Y}_{m},\,
\underbrace{c,\ldots,c}_{39-m}
\right]^{\top}.
\label{eq:pubsub_slice_state}
\end{equation}
Here, $X$ specifies the publisher state $x_0$, while $Y$ specifies the
common value of the first $m$ subscriber states. The remaining $39-m$
subscriber states are fixed at $c$ only when selecting the initial state at
which the value function is evaluated.

The corresponding two-dimensional slice is therefore
\begin{equation}
V_{m,c}(X,Y,\tau)
=
V\!\left(x^{m,c}(X,Y),\tau\right).
\label{eq:pubsub_value_slice}
\end{equation}
Thus, every point in a two-dimensional plot still corresponds to a complete
40-dimensional state. Importantly, Eq.~\eqref{eq:pubsub_slice_state} does
not constrain the subsequent trajectory to remain on this slice. Once the
initial state is specified, the value function is defined using the original
40-dimensional dynamics and all 39 control inputs. In other words, the
slice restricts only the set of initial states being visualized, not the
underlying dynamics.

Different choices of $(m,c)$ probe different portions of the
40-dimensional state space. Increasing $m$ changes how many subscriber
coordinates vary jointly with $Y$, while changing $c$ moves the slice along
the remaining subscriber directions. We therefore evaluate multiple
$(m,c)$ pairs rather than relying on a single two-dimensional cross-section.

\paragraph{Ground-truth value on the slices.}
A useful property of this benchmark is that the ground-truth reachability
value can be evaluated at an arbitrary 40-dimensional state without solving
a 40-dimensional gridded HJ equation. This follows from the particular
triangular structure of the publisher--subscriber dynamics.

Recall that
\[
\dot x_0 = a x_0,
\qquad
\dot x_i
=
-x_0+a x_i+b u_i-\gamma x_0^2x_i,
\qquad
i=1,\ldots,39.
\]
The publisher state $x_0$ is not affected by either the subscriber states or
the controls. Therefore, for an initial publisher state $x_0(0)$, its future
trajectory is known explicitly:
\begin{equation}
x_0(s)=x_0(0)e^{as}.
\label{eq:pubsub_x0_solution}
\end{equation}
Once this trajectory is fixed, each subscriber evolves independently
according to
\begin{equation}
\dot x_i
=
\left(a-\gamma x_0(s)^2\right)x_i
-x_0(s)+b u_i.
\label{eq:pubsub_scalar_ltv}
\end{equation}
Hence, conditioned on the publisher trajectory, the original 39 controlled
subscriber dynamics reduce to 39 independent scalar linear
time-varying systems.

To determine whether the target can be reached, consider first a fixed
candidate time $s\in[0,\tau]$. For Eq.~\eqref{eq:pubsub_scalar_ltv}, define
the scalar state-transition factor
\begin{equation}
\Phi(s,\sigma)
=
\exp\left(
\int_{\sigma}^{s}
\left(a-\gamma x_0(\rho)^2\right)d\rho
\right).
\label{eq:pubsub_transition}
\end{equation}
Using the variation-of-constants formula, the $i$th subscriber state at
time $s$ can be written as
\begin{equation}
x_i(s)
=
\bar{x}_i(s)
+
b\int_0^s
\Phi(s,\sigma)u_i(\sigma)\,d\sigma,
\label{eq:pubsub_subscriber_solution}
\end{equation}
where
\begin{equation}
\bar{x}_i(s)
=
\Phi(s,0)x_i(0)
-
\int_0^s
\Phi(s,\sigma)x_0(\sigma)\,d\sigma
\label{eq:pubsub_free_response}
\end{equation}
denotes the subscriber state that would result from the initial condition
and publisher forcing in the absence of control.

Since the controls satisfy
\[
|u_i(\sigma)|\le u_{\max},
\]
and $\Phi(s,\sigma)>0$, the control can move $x_i(s)$ within the interval
\begin{equation}
x_i(s)
\in
\left[
\bar{x}_i(s)-d(s),\,
\bar{x}_i(s)+d(s)
\right],
\label{eq:pubsub_reachable_interval}
\end{equation}
where
\begin{equation}
d(s)
=
b u_{\max}
\int_0^s
\Phi(s,\sigma)\,d\sigma
\label{eq:pubsub_control_radius}
\end{equation}
is the maximum displacement generated by the bounded input.

This interval characterization makes the optimal control at a fixed time
particularly simple. The target function is
\[
\ell(x)
=
\frac12
\left(
r^2-x_0^2-\frac{1}{39}\sum_{i=1}^{39}x_i^2
\right),
\]
so maximizing $\ell$ is equivalent to making the magnitudes of the
subscriber states as small as possible. If the interval in
Eq.~\eqref{eq:pubsub_reachable_interval} contains zero, the control can
drive the $i$th subscriber exactly to zero at time $s$. Otherwise, the
closest achievable state is the endpoint of the interval nearest to zero.
Consequently,
\begin{equation}
\min_{u_i(\cdot)}
|x_i(s)|
=
\left(
|\bar{x}_i(s)|-d(s)
\right)_+,
\label{eq:pubsub_min_subscriber}
\end{equation}
where $(z)_+=\max\{z,0\}$.

Because the 39 subscriber controls are independent, this minimization can
be performed independently for every subscriber. Therefore, the smallest
target quadratic term achievable at the fixed time $s$ is
\begin{equation}
q^\star(s;x)
=
x_0(s)^2
+
\frac{1}{39}
\sum_{i=1}^{39}
\left(
|\bar{x}_i(s)|-d(s)
\right)_+^2.
\label{eq:pubsub_qstar}
\end{equation}
The largest target-function value achievable exactly at time $s$ is then
\[
\ell^\star(s;x)
=
\frac12
\left(
r^2-q^\star(s;x)
\right).
\]

Finally, the reachability problem does not prescribe the exact time at which
the target must be entered; reaching the target at any time within
$[0,\tau]$ is sufficient. We therefore optimize over the candidate arrival
time $s$ as well. Under the target-positive convention used in this paper,
the ground-truth finite-horizon value is
\begin{equation}
V_{\mathrm{GT}}(x,\tau)
=
\max_{0\le s\le\tau}
\ell^\star(s;x)
=
\frac12
\left[
r^2-
\min_{0\le s\le\tau}
q^\star(s;x)
\right].
\label{eq:pubsub_gt}
\end{equation}
Thus, evaluating the ground truth does not require discretizing the full
40-dimensional state space. Instead, the special structure of the dynamics
reduces the calculation to scalar subscriber expressions together with a
one-dimensional optimization over the arrival time $s$. In implementation,
the required scalar integrals and the final time optimization are evaluated
numerically, yielding a semi-analytic ground-truth evaluator without a
high-dimensional HJ grid solve.

For each visualization slice, we use exactly the same 40-dimensional states
defined in Eq.~\eqref{eq:pubsub_slice_state} for both the learned value and
the ground truth:
\[
V\!\left(x^{m,c}(X,Y),\tau\right)
\qquad\text{and}\qquad
V_{\mathrm{GT}}\!\left(x^{m,c}(X,Y),\tau\right).
\]
We then extract and compare their zero-level contours in the $(X,Y)$ plane.
Therefore, the ground-truth curves shown in the two-dimensional figures are
not obtained from a separate two-dimensional HJ problem; they are
cross-sections of the ground-truth 40-dimensional reachable set boundary.
Evaluating several $(m,c)$ slices consequently provides a geometric
diagnostic of the learned 40-dimensional boundary across multiple portions
of the state space, while both training and ground-truth evaluation remain
defined on the original 40-dimensional system.

\newpage

\section{Additional Results and Ablation Studies}\label{app:result-details}

For each system, we visualize the value functions learned by Contract2Reach and MPC-DeepReach at every learning rate used in Table~\ref{tab:high_dimensional_lr_sensitivity} on six representative two-dimensional slices. The overlay figures superimpose the boundaries from all learning rates on each slice, so the spread of the curves directly shows the sensitivity to the learning rate. In the grid figures, each panel shows one slice (row) at one learning rate (column). All figures are collected at the end of this appendix.

\subsection{Parameterized 2D Drone: Extended Analysis}

Figures~\ref{fig:app_pvd_overlay}--\ref{fig:app_pvd_mpc_grid} show six slices of the PVD in the $(z,v)$ plane, each at a different value of the control gain $K$. For Contract2Reach, the boundaries learned with different learning rates are nearly indistinguishable from each other and from the reference on every slice. This holds even at $K=0.48$, where the gain is too small to counteract gravity and only a thin region of the state space avoids failure over the horizon. MPC-DeepReach is far more sensitive to the learning rate. With small learning rates, its boundary oscillates inside the safe set at the largest gains, and only the largest learning rate recovers the thin safe region at $K=0.48$, which the others miss entirely. Its accuracy improves steadily as the learning rate grows, so a good result requires tuning this parameter.

\subsection{13D Quadrotor: Extended Analysis}

Figures~\ref{fig:app_quad_overlay}--\ref{fig:app_quad_mpc_grid} show six slices in the $(p_x,p_y)$ plane around the cylindrical obstacle: hover, forward flight at three speeds, a $30^\circ$ roll, and a roll rate $\omega_x=2$. Since no reference solution exists for this system, these figures show how consistent each method is across learning rates rather than how accurate it is, and each panel reports the safe area of the slice instead of B-P95. For both methods, the boundaries learned with different learning rates nearly coincide on every slice, and the safe area changes by less than one percentage point. Both methods are therefore insensitive to the learning rate on this system, in agreement with Table~\ref{tab:high_dimensional_lr_sensitivity}.

\subsection{40D Publisher-Subscriber: Extended Analysis}

Figures~\ref{fig:app_pubsub_overlay}--\ref{fig:app_pubsub_mpc_grid} show six slices $x^{m,c}(X,Y)$ defined in Appendix~\ref{app:pubsub}, with the publisher state $x_0$ on the horizontal axis and the common value of the first $m$ subscriber states on the vertical axis. The two methods fail in different ways. Contract2Reach predicts a boundary on every slice at every learning rate, and its errors concentrate on slices whose reference set contains a narrow strip, such as $(m,c)=(3,-0.75)$. With small learning rates, it misses the strip and labels it as outside the certified set. This error is conservative, and the strip is recovered as the learning rate increases, although only partially on $(3,-0.75)$. MPC-DeepReach instead loses or distorts the boundary as a whole. On $(3,+1)$, it predicts no boundary at some learning rates and spurious boundaries near the edge of the domain at another. On $(12,-0.25)$, the reference boundary consists of four curves, but MPC-DeepReach recovers only two of them at every learning rate. The two curves it does recover lie close to the reference, so this error shows up in the curve count rather than in B-P95. Its accuracy also varies irregularly with the learning rate, so a small change in this parameter can turn an accurate boundary into a poor one.


\begin{center}
\includegraphics[width=\linewidth,height=0.4\textheight,keepaspectratio]{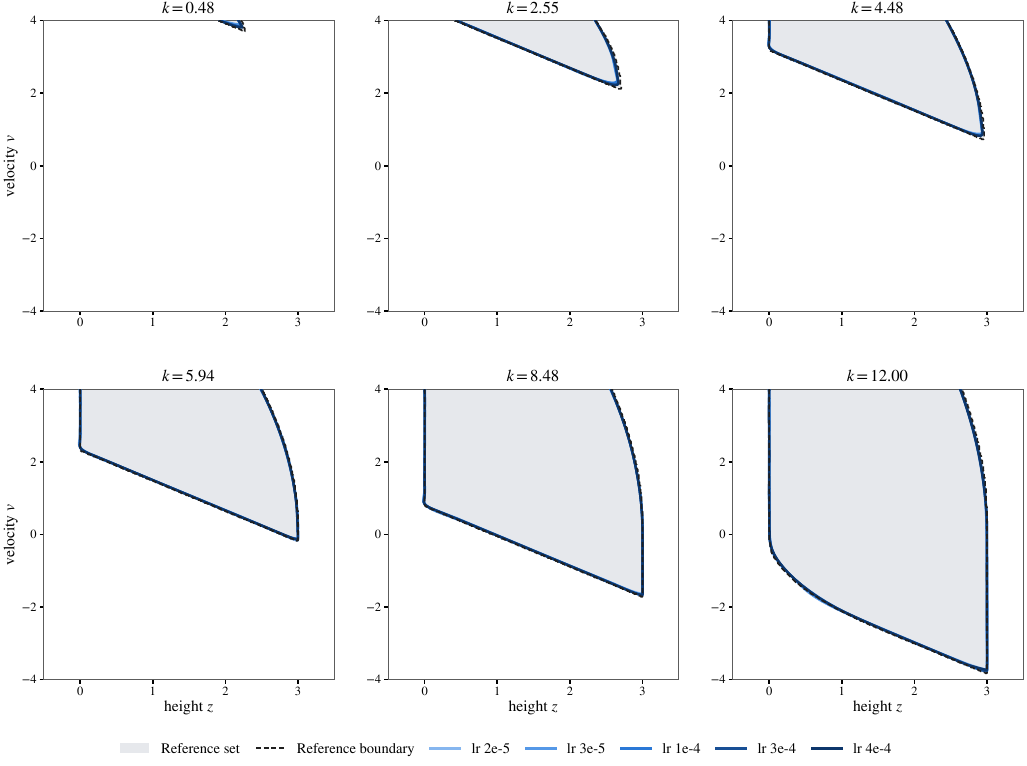}\\[6pt]
\includegraphics[width=\linewidth,height=0.4\textheight,keepaspectratio]{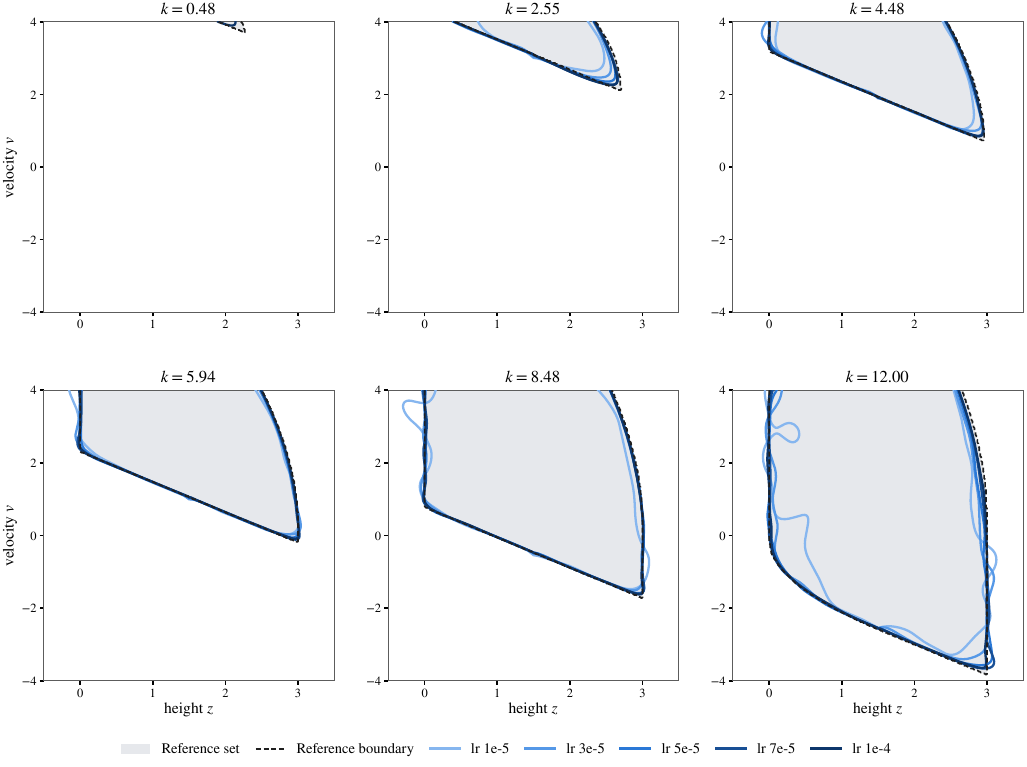}
\captionof{figure}{Boundaries for all learning rates superimposed on each PVD slice, for Contract2Reach (top) and MPC-DeepReach (bottom). The curves of Contract2Reach nearly coincide with the reference at every learning rate, whereas MPC-DeepReach shows spurious oscillations at small learning rates, most visibly at $K=8.48$ and $K=12$.}
\label{fig:app_pvd_overlay}
\end{center}

\newpage

\begin{center}
\includegraphics[width=\linewidth,height=0.8\textheight,keepaspectratio]{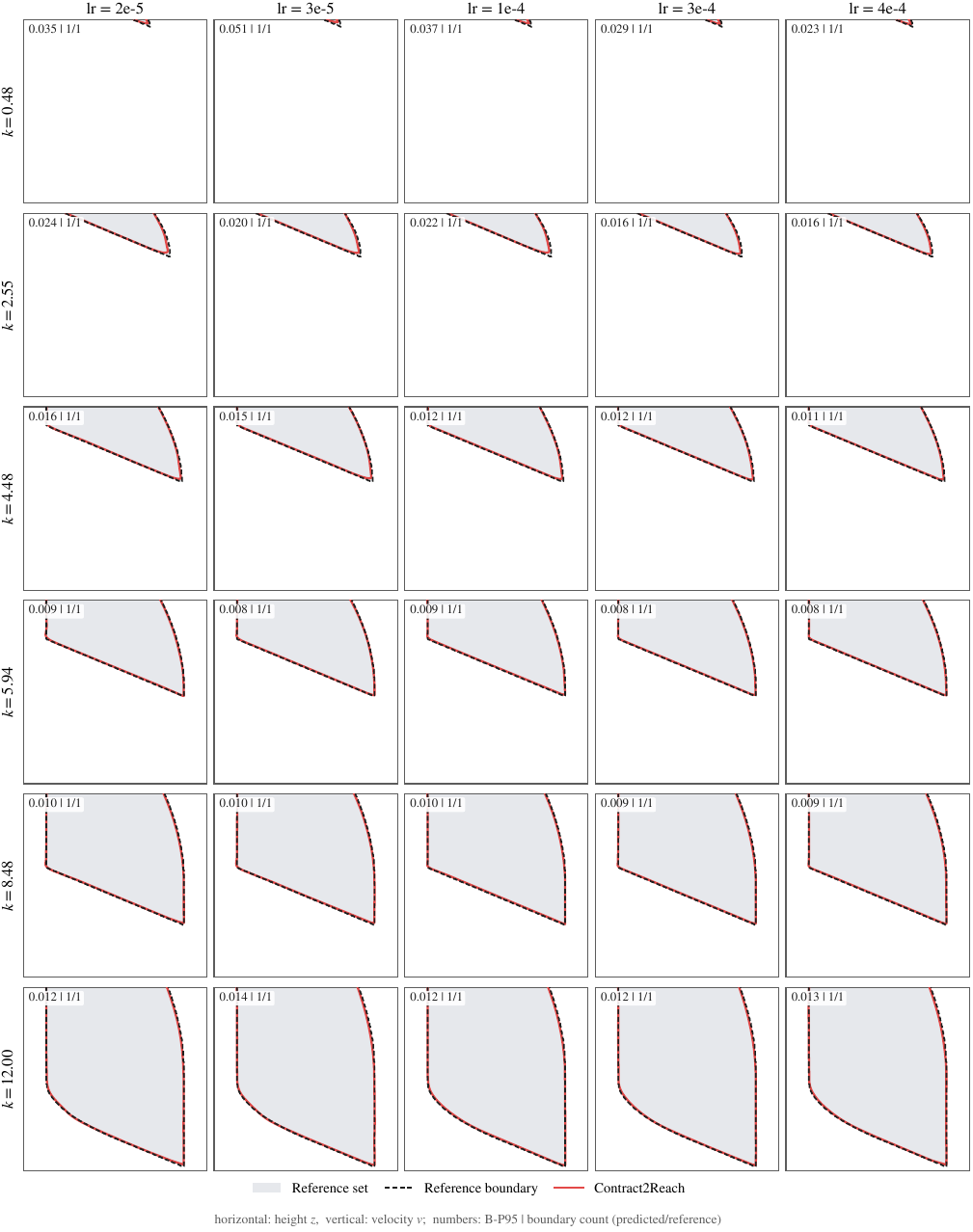}
\captionof{figure}{Contract2Reach on the PVD for each learning rate (columns) and slice $K$ (rows). Gray: reference set; dashed: reference boundary; solid: predicted boundary. Numbers: B-P95 $|$ predicted/reference boundary count.}
\label{fig:app_pvd_c2r_grid}
\end{center}

\newpage

\begin{center}
\includegraphics[width=\linewidth,height=0.8\textheight,keepaspectratio]{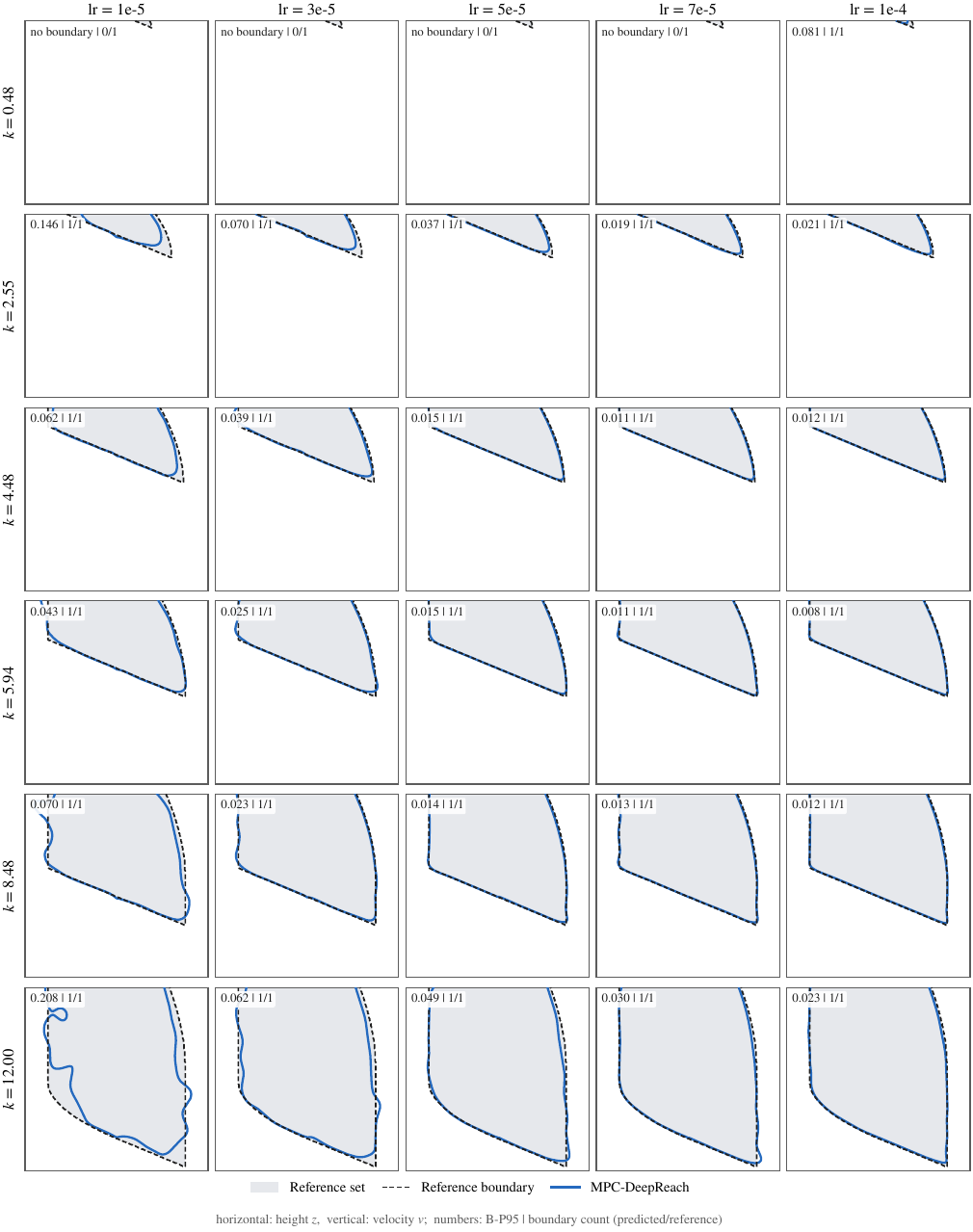}
\captionof{figure}{MPC-DeepReach on the PVD, in the same format as Figure.~\ref{fig:app_pvd_c2r_grid}. At $K=0.48$, no boundary is predicted at four of the five learning rates.}
\label{fig:app_pvd_mpc_grid}
\end{center}

\newpage

\begin{center}
\includegraphics[width=\linewidth,height=0.4\textheight,keepaspectratio]{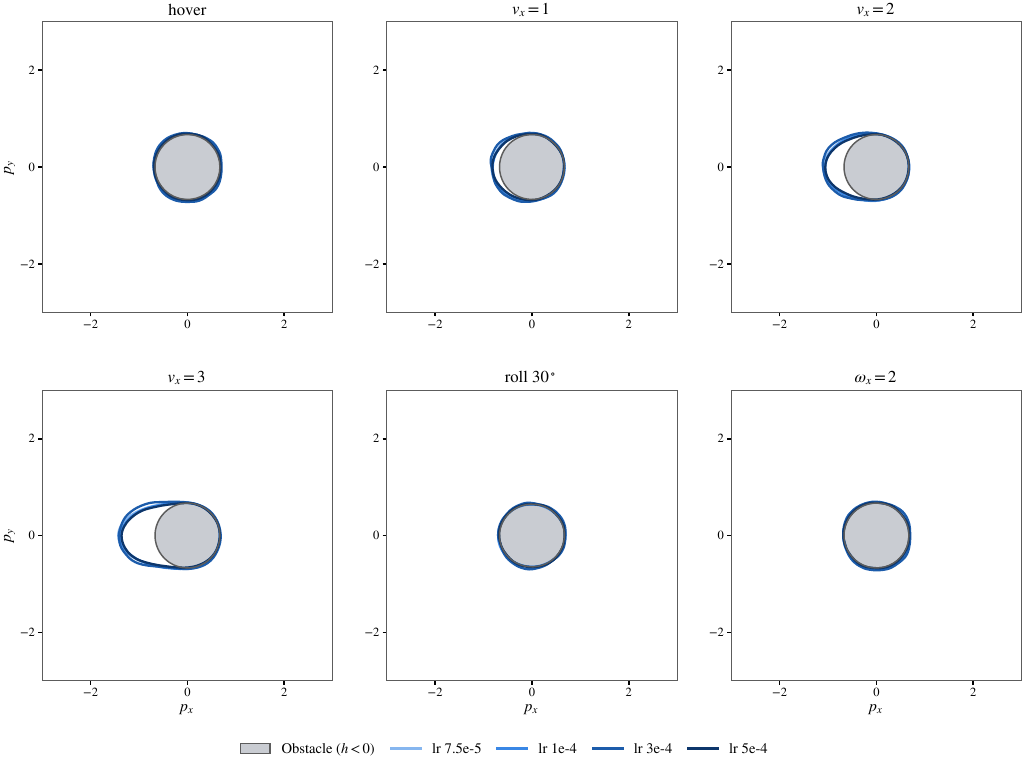}\\[6pt]
\includegraphics[width=\linewidth,height=0.4\textheight,keepaspectratio]{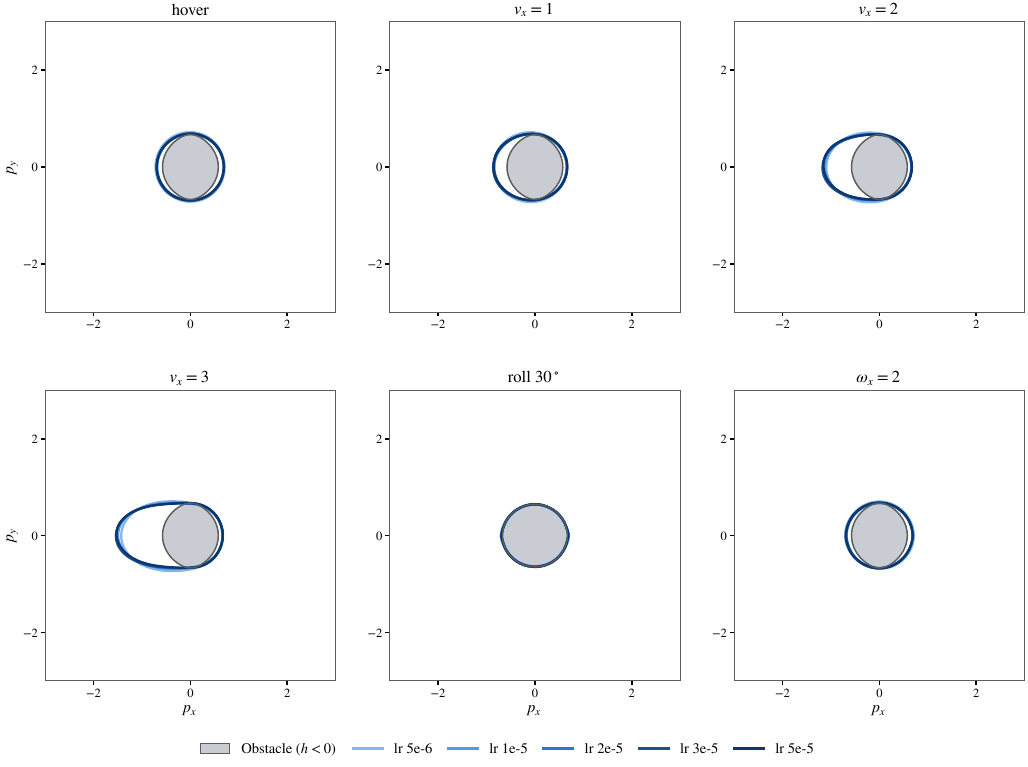}
\captionof{figure}{Boundaries for all learning rates superimposed on each Quadrotor 13D slice, for Contract2Reach (top) and MPC-DeepReach (bottom). For both methods, the curves nearly coincide across learning rates.}
\label{fig:app_quad_overlay}
\end{center}

\newpage

\begin{center}
\includegraphics[width=\linewidth,height=0.8\textheight,keepaspectratio]{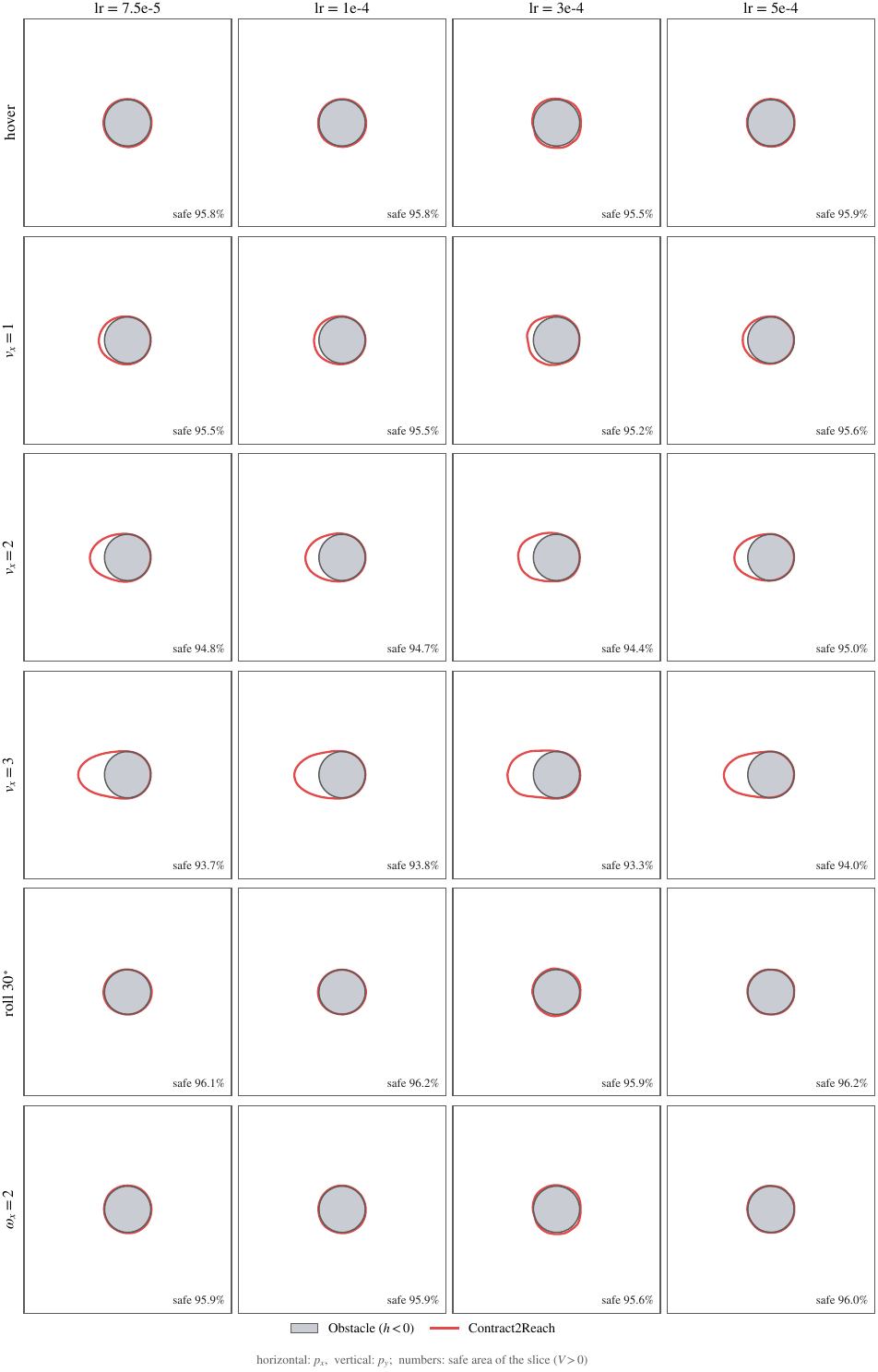}
\captionof{figure}{Contract2Reach on Quadrotor 13D for each learning rate (columns) and slice (rows). Gray: obstacle ($h<0$); solid: predicted boundary. Numbers: safe area of the slice ($V>0$). No reference solution is available.}
\label{fig:app_quad_c2r_grid}
\end{center}

\newpage

\begin{center}
\includegraphics[width=\linewidth,height=0.8\textheight,keepaspectratio]{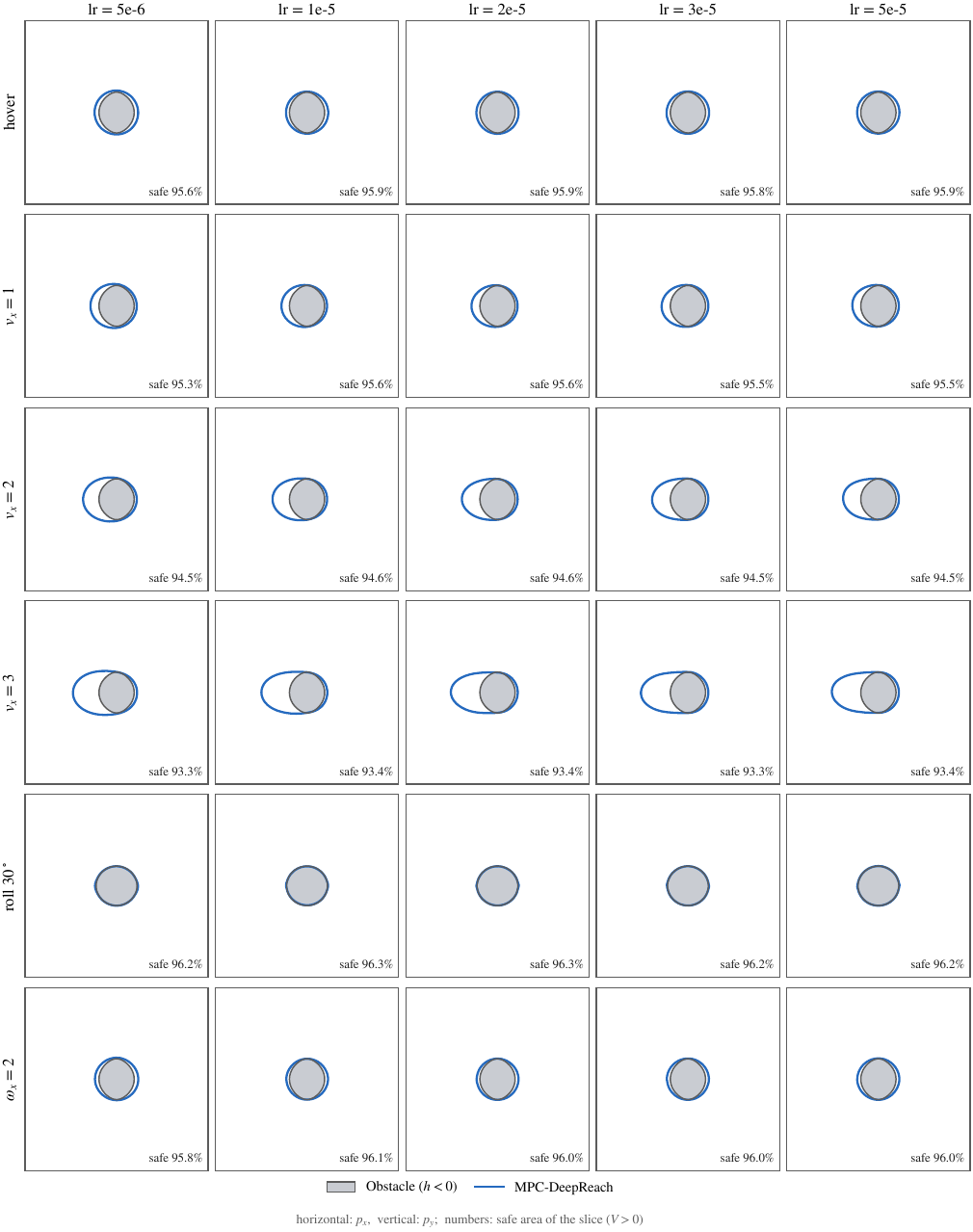}
\captionof{figure}{MPC-DeepReach on Quadrotor 13D, in the same format as Figure.~\ref{fig:app_quad_c2r_grid}.}
\label{fig:app_quad_mpc_grid}
\end{center}

\newpage

\begin{center}
\includegraphics[width=\linewidth,height=0.4\textheight,keepaspectratio]{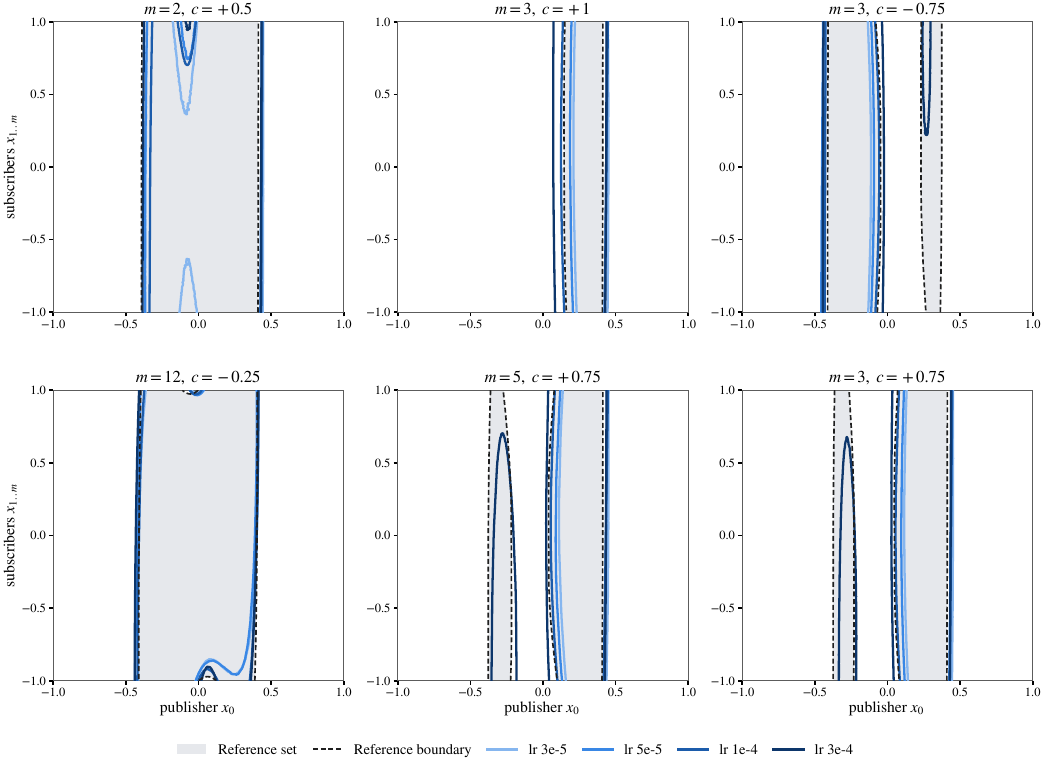}\\[6pt]
\includegraphics[width=\linewidth,height=0.4\textheight,keepaspectratio]{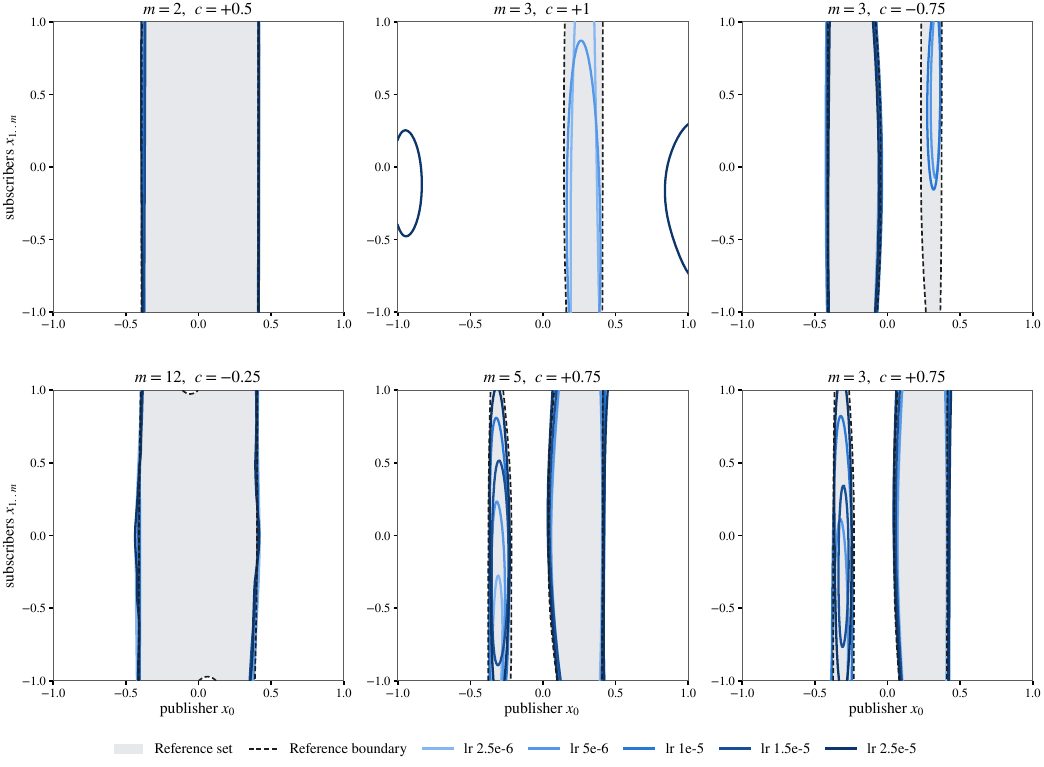}
\captionof{figure}{Boundaries for all learning rates superimposed on each PubSub 40D slice, for Contract2Reach (top) and MPC-DeepReach (bottom). Contract2Reach misses narrow strips of the reference set at smaller learning rates and approaches the reference as the learning rate increases, whereas MPC-DeepReach yields no boundary or spurious boundaries on $(m,c)=(3,+1)$ at some learning rates.}
\label{fig:app_pubsub_overlay}
\end{center}

\newpage

\begin{center}
\includegraphics[width=\linewidth,height=0.8\textheight,keepaspectratio]{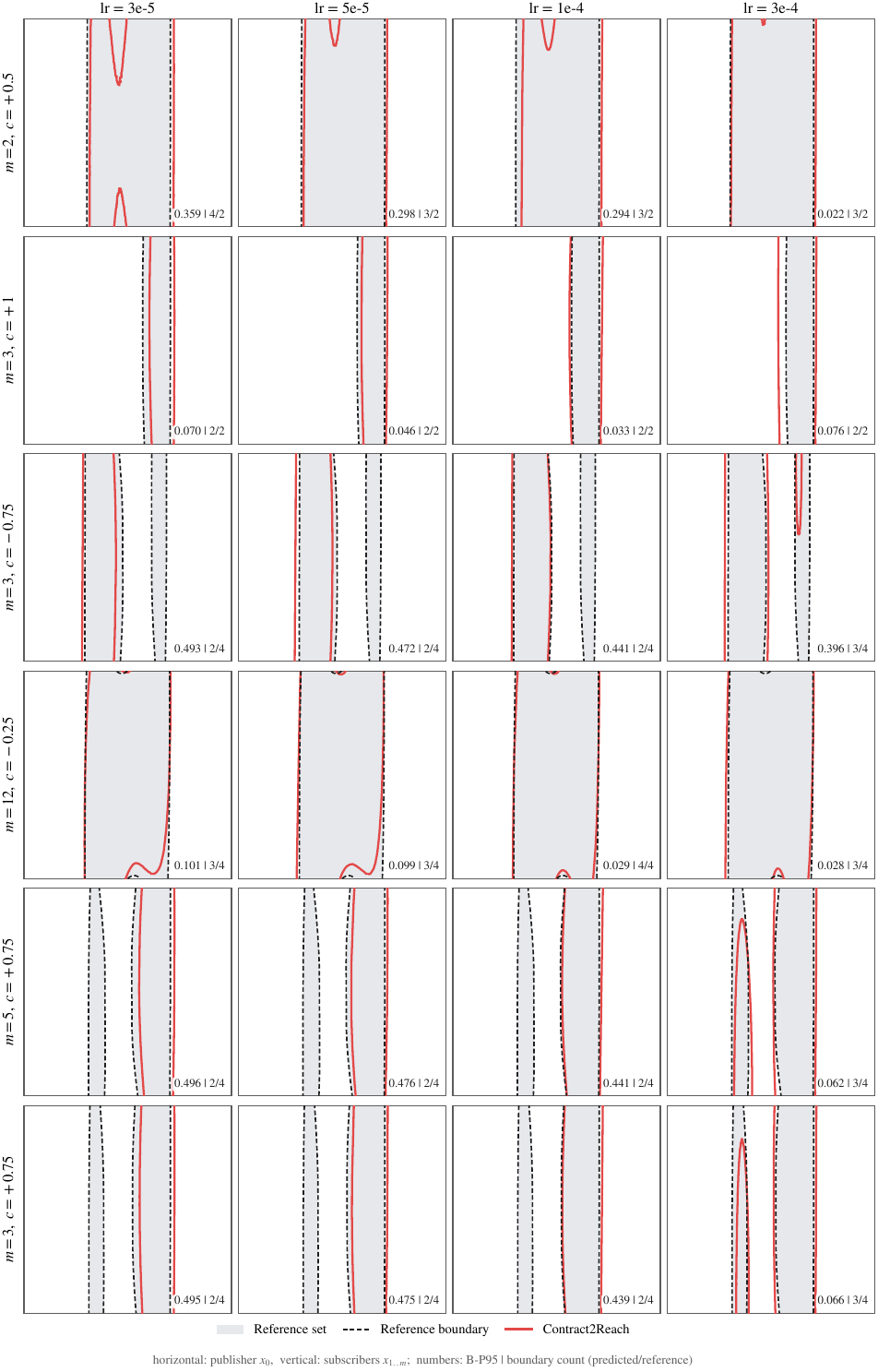}
\captionof{figure}{Contract2Reach on PubSub 40D for each learning rate (columns) and slice $(m,c)$ (rows). Gray: reference set; dashed: reference boundary; solid: predicted boundary. Numbers: B-P95 $|$ predicted/reference boundary count.}
\label{fig:app_pubsub_c2r_grid}
\end{center}

\newpage

\begin{center}
\includegraphics[width=\linewidth,height=0.8\textheight,keepaspectratio]{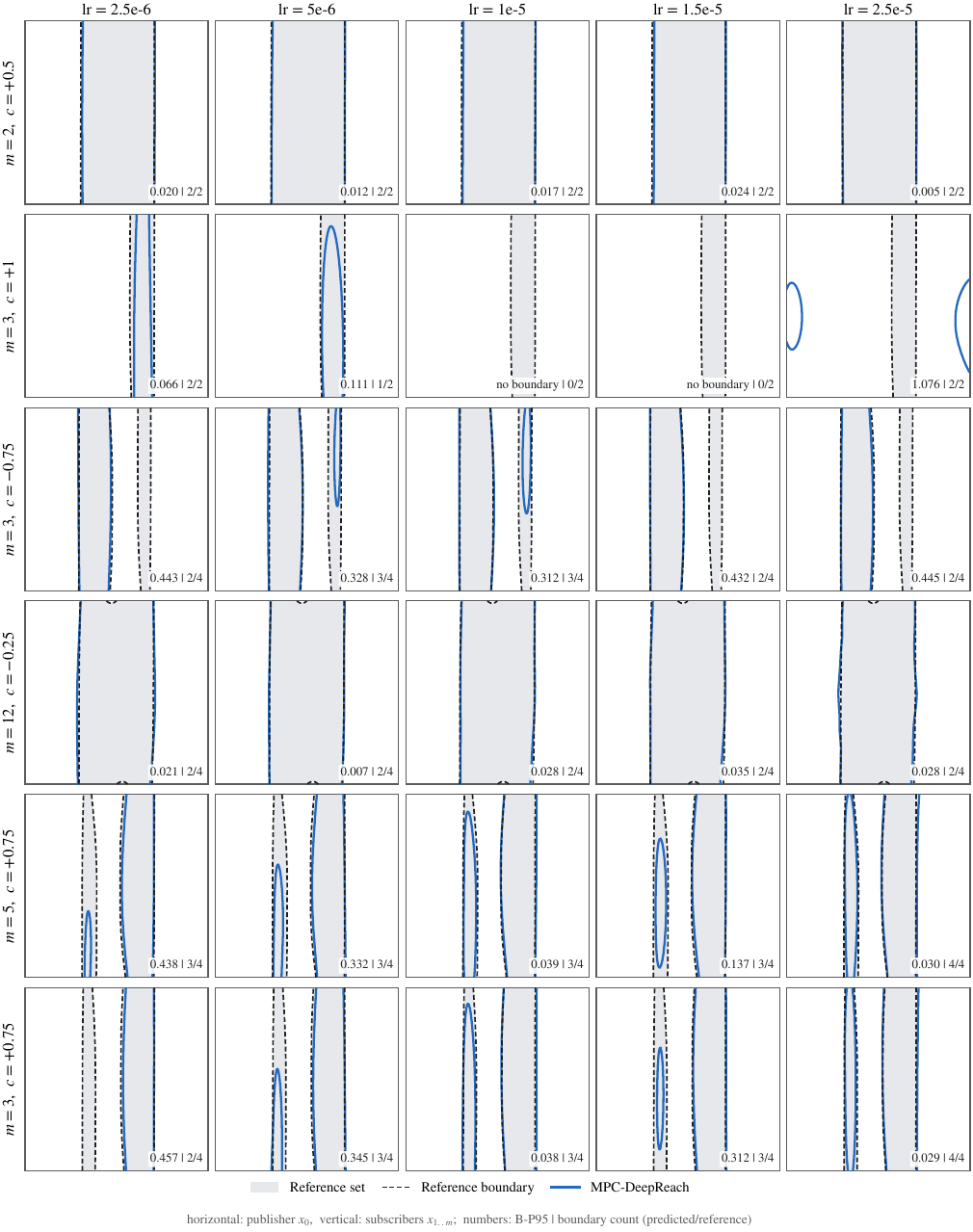}
\captionof{figure}{MPC-DeepReach on PubSub 40D, in the same format as Figure.~\ref{fig:app_pubsub_c2r_grid}. On the slice $(m,c)=(3,+1)$, no boundary is predicted at two learning rates.}
\label{fig:app_pubsub_mpc_grid}
\end{center}

\end{document}